\documentclass{aa}  

\newcommand{\hrieuv}{\text{HRI}_{\text{EUV}}}
\usepackage{graphicx}
\usepackage{bm}
\usepackage{natbib}
\usepackage{mathtools}
\usepackage{amsmath}
\usepackage{multirow}
\usepackage[export]{adjustbox}
\usepackage{rotating}

\usepackage{tikz}
\usepackage{tabularx}
\usepackage{txfonts}
\usepackage{textcomp}
\usepackage{color}

\begin{document} 

   \title{Accessing the fine temporal scale of EUV brightenings and their quasi-periodic pulsations: 1~second cadence observations by Solar Orbiter/EUI}
   
   \titlerunning{1~second cadence observations by Solar Orbiter/EUI}

   \author{%
            {Daye Lim}\inst{\ref{aff:ROB}, \ref{aff:CmPA}}
            {, Tom Van Doorsselaere}\inst{\ref{aff:CmPA}}
            {, Nancy Narang}\inst{\ref{aff:ROB}}
            {, Laura A. Hayes}\inst{\ref{aff:DIAS}}
            {, Emil Kraaikamp}\inst{\ref{aff:ROB}}
            {, Aadish Joshi}\inst{\ref{aff:CmPA}}
            {, Konstantina Loumou}\inst{\ref{aff:ROB}}
            {, Cis Verbeeck}\inst{\ref{aff:ROB}}
            {, David Berghmans}\inst{\ref{aff:ROB}}
            \and
            {Krzysztof Barczynski}\inst{\ref{aff:ETH}, \ref{aff:Davos}}
          }
 
   \institute{%
             \label{aff:ROB}{Solar-Terrestrial Centre of Excellence – SIDC, Royal Observatory of Belgium, Ringlaan -3- Av. Circulaire, 1180 Brussels, Belgium} \email{daye.lim@oma.be}
            \and
             \label{aff:CmPA}{Centre for mathematical Plasma Astrophysics, Department of Mathematics, KU Leuven, Celestijnenlaan 200B, 3001 Leuven, Belgium} 
            \and
            \label{aff:DIAS}{Astronomy \& Astrophysics Section, School of Cosmic Physics, Dublin Institute for Advanced Studies, DIAS Dunsink Observatory, Dublin, D15 XR2R, Ireland.}
            \and
            \label{aff:ETH}{ETH-Zurich, Wolfgang-Pauli-Str. 27, 8093 Zurich, Switzerland}
            \and
            \label{aff:Davos}{Physikalisch-Meteorologisches Observatorium Davos, World Radiation Center, 7260, Davos Dorf, Switzerland}
             }

   \authorrunning{Lim et al.}

  \abstract
   {Small-scale extreme-ultraviolet (EUV) transient brightenings are observationally abundant and critically important to investigate.  Determining whether they share the same physical mechanisms as larger-scale flares would have significant implications for the coronal heating problem. A recent study has revealed that quasi-periodic pulsations (QPPs), a common feature in both solar and stellar flares, may also be present in EUV brightenings in the quiet Sun (QS).}
   {We aim to characterise the properties of EUV brightenings and their associated QPPs in both QS and active regions (ARs) using unprecedented 1~s cadence observations from Solar Orbiter’s Extreme Ultraviolet Imager (EUI).}
   {We applied an automated detection algorithm to analyse statistical properties of EUV brightenings. QPPs were identified using complementary techniques optimised for both stationary and non-stationary signals, including a Fourier-based method, ensemble empirical mode decomposition, and wavelet analysis.}
   {Over 500~000 and 300~000 brightenings were detected in ARs and QS regions, respectively. Brightenings with lifetimes shorter than 3~s were detected, demonstrating the importance of high temporal resolution. The QPP occurrence rates were approximately 11\% in AR brightenings and 9\% in QS brightenings, with non-stationary QPPs being more common than stationary ones. QPP periods span from 5 to over 500~s and show similar distributions between AR and QS. Moderate linear correlations were found between QPP periods and the lifetime and spatial scale of the associated brightenings, while no significant correlation was found with peak brightness. We found a consistent power-law scaling, with a weak correlation and a large spread, between QPP period and lifetime in EUV brightenings, solar, and stellar flares.}
   {The results support the interpretation that EUV brightenings may represent a small-scale manifestation of the same physical mechanisms driving larger solar and stellar flares. Furthermore, the similarity in the statistical properties of EUV brightenings and their associated QPPs between AR and QS regions suggests that the underlying generation mechanisms may not strongly depend on the large scale magnetic environment.}

   \keywords{Sun: UV radiation --
             Sun: atmosphere --
             Sun: corona --
             Sun: oscillations --
             Waves --
             Stars: oscillations (including pulsations)
            }

\maketitle

\section{Introduction}

The coronal heating problem remains one of the most persistent and fundamental challenges in astrophysics \citep{2015RSPTA.37340256K}. Among the various mechanisms proposed to address this issue, the nanoflare heating theory has attracted considerable attention. This theory posits that the solar corona may be heated by a sufficiently large number of small-scale energy release events \citep{1988ApJ...330..474P}. Imaging observations in the extreme-ultraviolet (EUV) wavelengths have consistently revealed ubiquitous small, transient brightenings in the quiet Sun (QS) corona, which may correspond to such small-scale flaring events. Although these features have been referred to by various names depending on the observational instruments and analysis methods \citep{2002ApJ...572.1048A, 2003A&A...409..755H, 2021A&A...656L...4B, 2022A&A...661A.149P, 2023A&A...671A..64D}, in this study, we refer to them collectively as EUV brightenings.

As the spatial resolution of solar imaging instruments has progressively improved, the record for the smallest observable events has been repeatedly updated. A statistically significant number of EUV brightenings were first detected by the Solar and Heliospheric Observatory’s Extreme ultraviolet Imaging Telescope (SOHO/EIT; \citealt{1995SoPh..162..291D}) at 195~\AA\, which revealed events spanning 10–300 $\text{Mm}^2$ \citep{1998A&A...336.1039B}. Subsequent observations with the Transition Region and Coronal Explorer (TRACE; \citealt{1999SoPh..187..229H}) at 173 and 195~\AA\,, the Atmospheric Imaging Assembly (AIA; \citealt{2012SoPh..275...17L}) at 171~\AA\ as well as other coronal passbands (e.g. 193, 211, 94, and 131~\AA) onboard the Solar Dynamics Observatory (SDO), and the High-resolution Coronal imager (Hi-C; \citealt{2013Natur.493..501C}) at 193~\AA\ have pushed this limit further, revealing brightenings as small as 0.7 \citep{2000ApJ...529..554P} and 0.2~$\text{Mm}^2$ \citep{2021A&A...647A.159C, 2017A&A...599A.137B}, respectively. This trend culminated in the launch of Solar Orbiter \citep{2020A&A...642A...1M} and its Extreme Ultraviolet Imager (EUI; \citealt{2020A&A...642A...8R}). The recent perihelion observations by EUI's High Resolution Imager 174~\AA\ ($\hrieuv$) revealed brightenings corresponding to a spatial scale of approximately 105~km, the smallest EUV brightenings detected to date in QS \citep{2025A&A...699A.138N}. Some of the EUI brightenings observed with EUI $\hrieuv$ may also correspond to small-scale coronal jets \citep{2021ApJ...918L..20H, 2023Sci...381..867C}.

While EUV brightenings are commonly observed in the QS, they have also been reported in active regions (ARs). For instance, SOHO/EIT 195~\AA\ observations revealed EUV brightenings in ARs with sizes ranging from approximately 6 to 500~$\text{Mm}^2$, occurring at a frequency of about one event every 10 seconds \citep{1999SoPh..186..207B} which is higher than the 0.06 events per second reported in the QS with the same instrument \citep{1998A&A...336.1039B}. Although subsequent instruments with improved spatial resolution, such as TRACE \citep{2001ApJ...563L.173S}, Hi-C \citep{2013ApJ...770L...1T, 2014ApJ...784..134R, 2016ApJ...822...35A, 2018A&A...615A..47S} and AIA \citep{2014ApJ...783...12U}, have occasionally detected similar events in ARs, they have not been studied as systematically as those in the QS. Moreover, despite its superior spatial resolution, EUI has not yet been employed to investigate EUV brightenings in ARs, leaving open questions about how a more complex magnetic environment influences these phenomena.

A central question in the nanoflare heating scenario is whether the flare energy distribution is steep enough for the numerous weakest events to contribute significantly to coronal heating \citep{1991SoPh..133..357H}. Addressing this requires not only identifying smaller-scale events but also uncovering those that remain hidden due to limited temporal resolution. Indeed, as imaging cadence has improved, progressively shorter-lived EUV brightenings have been detected \citep{2021A&A...647A.159C, 2021A&A...656L...4B}. For example, using EUI data with a 3~s cadence, event lifetimes were found to reach this lower bound, and their distribution followed a power law, implying many more undetected events at even shorter timescales \citep{2025A&A...699A.138N}. The EUI not only provides the highest spatial resolution to date but also allows for unprecedented temporal resolution. In this study, we utilise EUI $\hrieuv$ observations with a 1~s cadence to explore this previously inaccessible temporal regime, investigating short-lived EUV brightenings in both QS and ARs.

Exploring the abundance of small-scale EUV brightenings, alongside efforts to unveil the physical mechanisms driving these events, is of great importance. Evidence from 3D radiative MHD simulations \citep{2021A&A...656L...7C} and magnetic extrapolation \citep{2022SoPh..297..141B} suggested that most EUV brightenings are driven by magnetic reconnection. Follow-up observational studies reported that many events occur near magnetic neutral lines in bipolar regions, further supporting their reconnection origin \citep{2021ApJ...921L..20P, 2022A&A...660A.143K}. However, a larger statistical analysis by \citet{2024A&A...692A.236N} showed that only 12\% of EUV brightenings occur in clearly bipolar regions, highlighting not only the complexity of their magnetic environments but also the need for higher-resolution magnetic context \citep{2023ApJ...958L..38N}. Alternatively, impulsive Alfvén waves in the chromosphere have been suggested as a possible driver of EUV brightenings, based on recent simulation results \citep{2024ApJ...969L..34K}.  

Recent statistical analysis of EUV brightenings in the QS revealed that a subset of these events exhibits quasi-periodic pulsations (QPPs) \citep{2025A&A...698A..65L}, an intrinsic feature of solar and stellar flares \citep{2009SSRv..149..119N, 2021SSRv..217...66Z}. The QPPs showed periodicities ranging from approximately 15 to 260 seconds, with occurrence rates increasing with event brightness, consistent with observations in solar flares \citep{2020ApJ...895...50H}. In agreement with previous findings of QPPs in solar \citep{2016ApJ...833..284I, 2017A&A...608A.101P, 2018SoPh..293...61D, 2020ApJ...895...50H} and stellar flares \citep{2016MNRAS.459.3659P, 2025A&A...700A.178J}, no significant correlation was identified between the QPP period and peak brightness. However, in contrast to trends observed in solar flares \citep{2019A&A...624A..65P, 2020ApJ...895...50H}, the QPP period was found to be independent of either the event lifetime or length scale. The results demonstrated the robust presence of QPPs in EUV brightenings, supporting the interpretation that EUV brightenings may undergo processes similar to those of standard flares. 

While these studies have provided an initial understanding of QPPs observed in EUV brightenings, they have been limited to stationary features. Extending the analysis to include non-stationary QPPs, which constitute the majority of observed QPP events \citep{2019PPCF...61a4024N, 2023MNRAS.523.3689M}, would allow for a more comprehensive and statistically robust characterisation of QPP properties. Furthermore, expanding the investigation to ARs will provide insight into how the occurrence and characteristics of QPPs vary across different magnetic regions, thereby deepening our understanding of the underlying physical processes. In this study, we investigate EUV brightenings and their QPPs in both QS and ARs. Furthermore, by taking advantage of EUI $\hrieuv$ observations with a 1~s cadence, we are able to detect events with even shorter lifetimes and QPPs with shorter periods than previously possible. 

\section{Solar Orbiter/EUI $\hrieuv$}\label{sec:data}

\begin{figure*}
  \resizebox{\hsize}{!}{\includegraphics{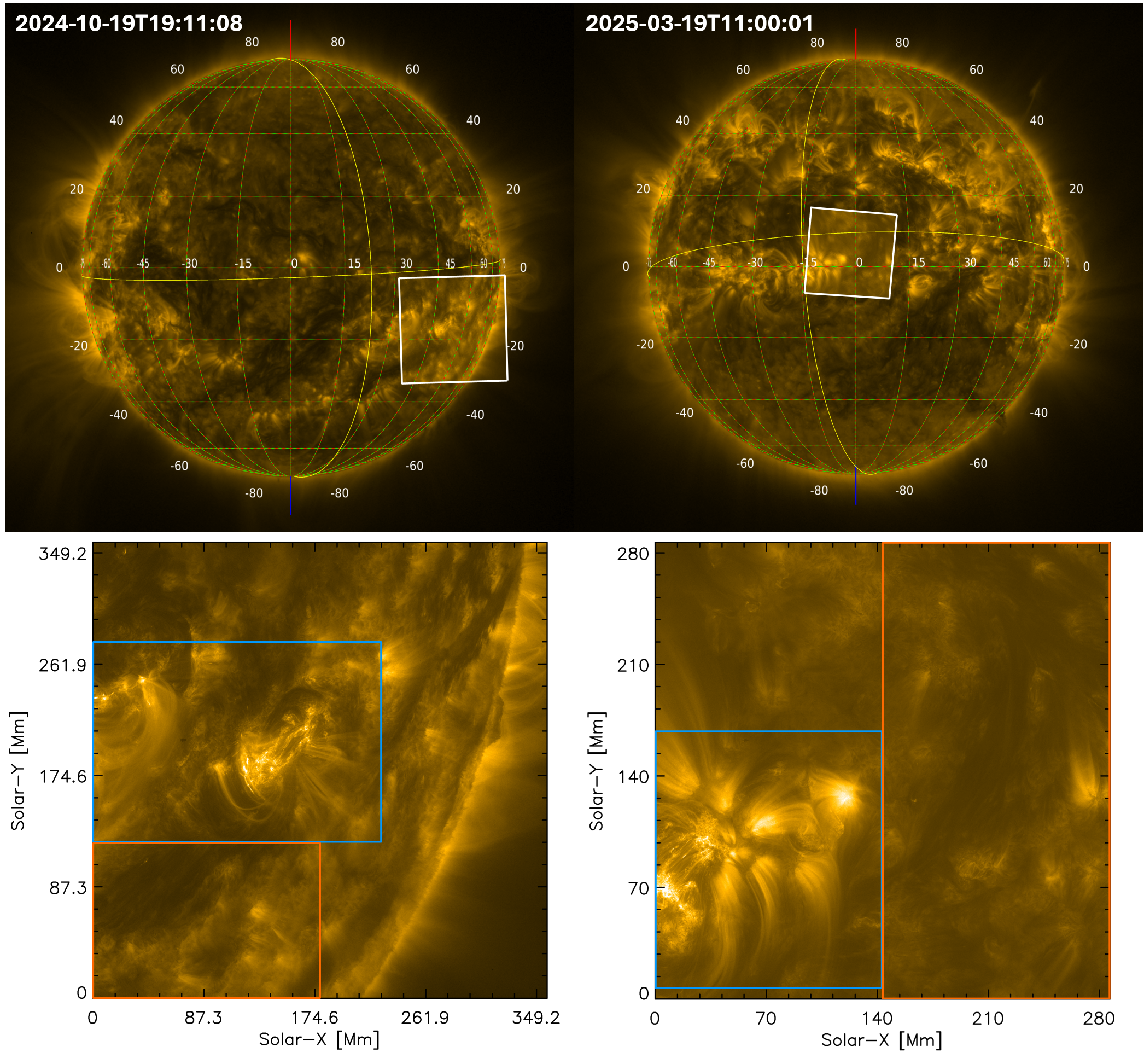}}
  \caption{Representative $\hrieuv$ 174~\AA\ images. Images were acquired by Solar Orbiter/EUI on 19 October 2024 at 19:11:08 UT (left) and 19 March 2025 at 11:00:01 UT (right). Top panels: Visualisations from JHelioviewer \citep{2017A&A...606A..10M}, showing $\hrieuv$ images (white rectangles) overlaid with near-simultaneous FSI images. Bottom panels: Full field-of-view $\hrieuv$ images, with blue and orange rectangles marking the active region and quiet Sun regions analysed in this study.}
  \label{fig:obs}
\end{figure*}

The EUI $\hrieuv$ was operated on 19 October 2024 at a fast imaging cadence of 1~s, marking the first instance of $\hrieuv$ acquiring images at this high cadence for scientific purposes. This observation was executed as part of a campaign titled R\_SMALL\_HRES\_HCAD\_RS-burst\footnote{\url{https://s2e2.cosmos.esa.int/confluence/display/SOSP/R_SMALL_HRES_HCAD_RS-burst}} Solar Orbiter Observing Plan (SOOP; \citealt{2020A&A...642A...3Z, 2020A&A...642A...6A}). This campaign was carried out between 19:04:01 and 19:32:00 UT, with each image taken at an exposure time of 0.659~s. During this period, the corona in 174~\AA\ was successfully imaged, except for intervals when the EUI Full Sun Imager (FSI) was observing simultaneously. Taking this into account, the longest uninterrupted sequence was recorded between 19:11:08 and 19:20:12 UT, comprising a total of 545 images. At the time, the Solar Orbiter was located at a distance of approximately 0.49 au from the Sun, corresponding to a pixel plate scale of around 174~km. A representative snapshot from this dataset is shown in the left panel of Fig.~\ref{fig:obs}.

Encouraged by the success of this campaign, a subsequent 1~s cadence (0.659~s exposure time) observation was performed on 19 March 2025, between 11:00:01 and 11:28:01 UT. In this run, the FSI was paused throughout the period, enabling the acquisition of a continuous sequence consisting of 1~681 images at 1~s intervals. During this observation, the Solar Orbiter was closer to the Sun, at a distance of approximately 0.39~au, resulting in an improved pixel scale of about 140~km. A representative example from this sequence is presented in the right panel of Fig.~\ref{fig:obs}.

As shown in Fig.~\ref{fig:obs}, each observation captures both active regions (ARs) and relatively quiet Sun (QS) regions within the same field of view (FOV). The sequence observed on 19 October 2024 includes NOAA ARs 13859 and 13860, while the sequence from 19 March 2025 encompasses NOAA ARs 14028 and 14029. For this study, we used calibrated level-2 data \citep{euidatarelease6}, and the analysis was restricted to the AR (indicated by the blue rectangle) and QS (the orange rectangle) subregions within the full FOV of each $\hrieuv$ observation. We define QS regions as areas that are not classified as NOAA ARs and that exhibit relatively weak photospheric magnetic fields. Both the AR and QS datasets possess identical spatial and temporal resolution. This allows us to minimise the influence of observational biases and attribute any differences in the measured properties primarily to the underlying differences in magnetic activity between the regions.

We would like to note that $\hrieuv$ sequences include spacecraft jitter. A detection scheme for EUV brightenings introduced in Section \ref{subsec:detectionscheme} was applied to the sequence of Carrington projected images, which generally ensures good alignment within each sequence. However, inspection of our dataset revealed that a residual jitter of up to approximately one pixel was still present. To assess whether such residual jitter could affect the detection results, we tested one of the four datasets, the AR dataset from 19 March 2025, by additionally applying a cross-correlation technique \citep{2022A&A...667A.166C} to minimise jitter. This reduced the residual jitter to about 0.1~pixel. Applying the same detection scheme to this corrected sequence and comparing the resulting event histograms with those shown in Fig.~\ref{fig:hist_campfires} demonstrated that the residual jitter had no measurable effect on the detection outcomes. Additionally, since a remaining jitter of up to one pixel is not expected to have any meaningful impact on the integrated light curves (described in Section \ref{sec:qpp}) considered for the QPP analysis, we therefore considered the Carrington projected sequence sufficient for all analyses in this study. 


\section{EUV brightenings}\label{sec:campfires}

\subsection{Analysis}\label{subsec:detectionscheme}

The detection of small-scale EUV brightenings in high-resolution $\hrieuv$ data has been carried out using the same automated detection algorithm\footnote{\url{https://github.com/frederic-auchere/campfires}} as in previous studies \citep{2023A&A...671A..64D, 2024A&A...688A..77D, 2023A&A...676A..64N, 2024A&A...692A.236N, 2025A&A...699A.138N}, to ensure methodological consistency. Originally introduced by \citet{2021A&A...656L...4B}, this technique identifies EUV brightenings based on significant coefficients in the first two spatial scales of an à trous wavelet transform, employing a B3-spline scaling function. A coefficient is deemed significant when its amplitude exceeds $n$-times the root-mean-square value expected from the instrument’s noise characteristics, accounting for both photon shot noise and detector read noise \citep{2023arXiv230714182G}. The code was further optimised for $\hrieuv$ science-phase data by \citet{2025A&A...699A.138N}, and we adopted this latest version in the present study. A comprehensive description of the methodology is provided in Appendix B of \citet{2021A&A...656L...4B} and Appendix A of \citet{2025A&A...699A.138N}. 

To determine an appropriate threshold value $n$ for distinguishing genuine EUV brightenings from noise, we applied the detection algorithm to the first image of each dataset while varying $n$ from 2 to 15. This approach was based on the assumption that the optimal value of $n$ would remain consistent across frames within a given continuous sequence. Moreover, this strategy was adopted to reduce computational cost, as applying the algorithm to full sequences, particularly with lower $n$ values, requires substantial processing time due to the large number of detected events. For each value, the total number of detected events was recorded. We then followed the elbow method\footnote{\url{https://pypi.org/project/kneed/}}, commonly used in $k$-means clustering, to identify the value of $n$ beyond which the number of detections does not decrease significantly compared to lower thresholds \citep{2020A&A...644A.152E}. This analysis yielded an optimal $n$ of 4 for the QS dataset from 19 March 2025, and 5 for the remaining three datasets. The results indicate that the optimal $n$ does not vary strongly between AR and QS regions. For reference, \citet{2025A&A...699A.138N} employed $n = 6$ for QS regions, and to enable a direct comparison with the EUV brightening birthrate derived in this study (see Section \ref{sec:birthrate} and Appendix \ref{appendix:influence_threshold}), we also ran our code for the QS dataset using the same threshold of 6. A comparison of the parameter histograms obtained with different thresholds shows that, apart from the overall decrease in the number of detected events at higher thresholds, the shapes of the distributions remain essentially unchanged. This demonstrates that our detection results, except for the birthrates, are not sensitive to the exact choice of threshold.

\subsection{Results and discussion}
\begin{figure*}
  \resizebox{\hsize}{!}{\includegraphics{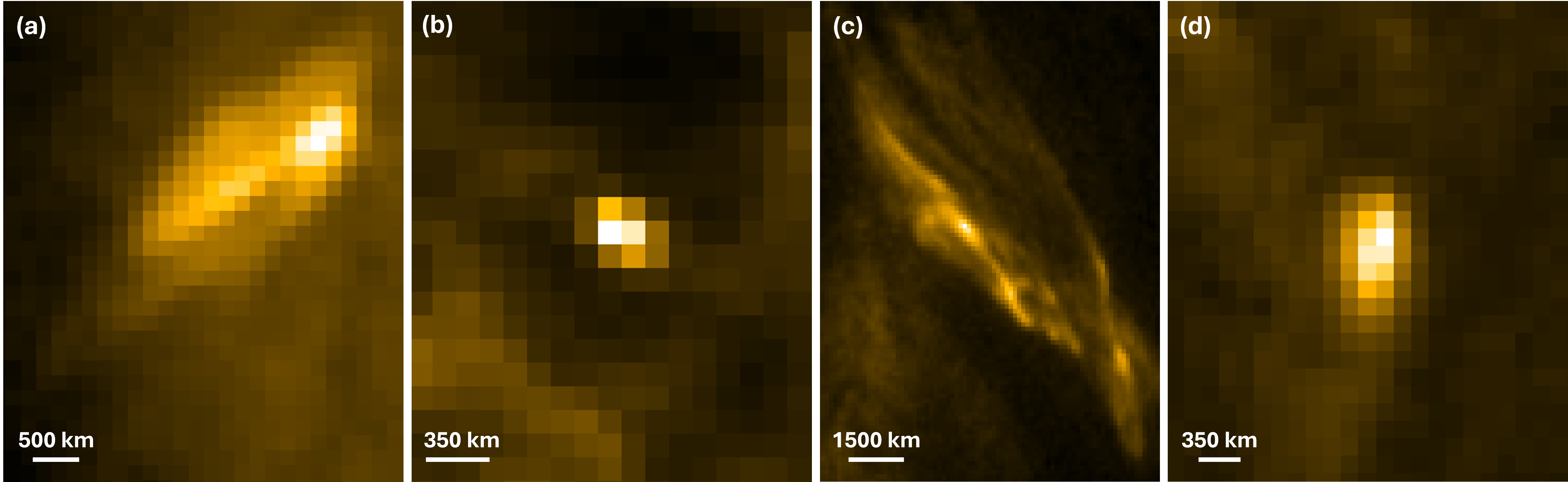}}
  \caption{Examples of the detected EUV brightenings. Panels (a) and (b) show events observed in an active region on 19 October 2024, while panels (c) and (d) show events observed in the quiet Sun on 19 March 2025. In each case, the image corresponds to the time of peak brightness of the event.}
  \label{fig:campfires}
\end{figure*}

Applying the detection algorithm to the full sequence of each dataset, using the respective optimal threshold values of $n$, resulted in the detection of the following numbers of EUV brightenings: 185 169 from the AR on 19 October 2024, 42 971 from the QS on 19 October 2024, 675 730 from the AR on 19 March 2025, and 474 345 from the QS on 19 March 2025. To examine whether EUV brightenings exhibit any preferential morphology in ARs or QS regions, we compared representative examples from each dataset. As illustrated in Fig.~\ref{fig:campfires}, no significant morphological differences were identified between the two solar regions. Additional examples are provided in Appendix \ref{appendix:example}. Although the two datasets from 19 October 2024 and 19 March 2025 differ in projection and exhibit minor variation in their individual results, these differences are not significant enough to affect the main statistical results or the overarching conclusions. Therefore, to focus on the distinctions between AR and QS regions rather than on dataset-specific effects, we present combined results for most analyses, with the exception of birthrate.

\subsubsection{Histograms of EUV brightening characteristics}

\begin{figure*}
  \resizebox{\hsize}{!}{\includegraphics{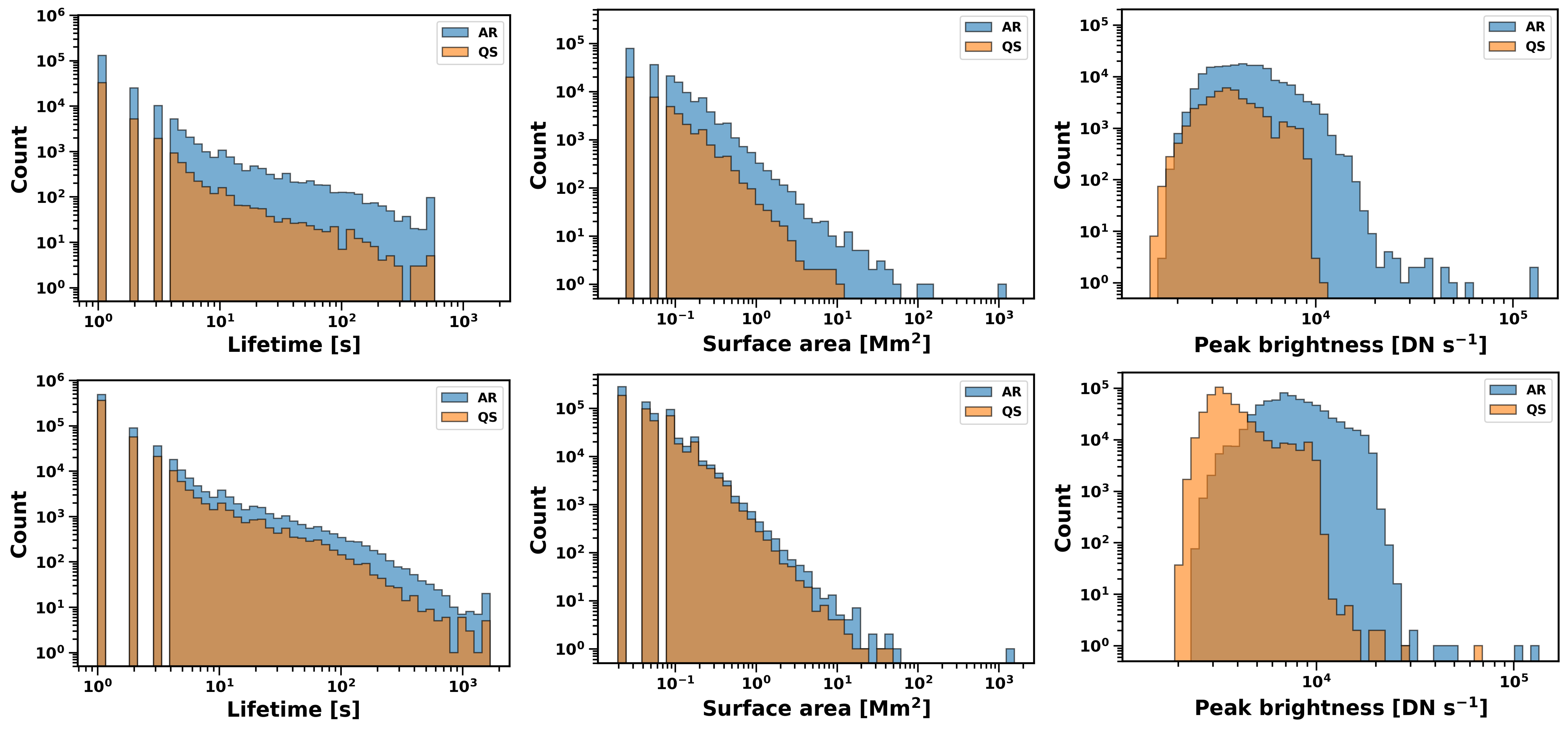}}
  \caption{Logarithmic histograms of EUV brightening properties. The panels show the lifetime (left), surface area (middle), and peak brightness (right) for events detected on 19 October 2024 (top) and 19 March 2025 (bottom). Events detected in active regions and the quiet Sun regions are shown in blue and orange, respectively.} 
  \label{fig:hist_campfires}
\end{figure*}

Fig.~\ref{fig:hist_campfires} presents histograms of the lifetime, surface area, and peak brightness of the detected events. The lifetime of an event is its duration in seconds. The surface area of an event is defined as the total area, in $\text{Mm}^{2}$, of the union of all pixels in the image plane that constitute the projection of the event over its entire lifetime. The peak brightness, expressed in DN $\text{s}^{-1}$, is the data value of the brightest voxel of the event. The descriptions of these event properties are provided in Appendix A of \citet{2025A&A...699A.138N}. The distributions show that both the surface area and lifetime extend down to the observational limits imposed by the spatial (0.02~$\text{Mm}^{2}$ on 19 October 2024 and 0.02~$\text{Mm}^{2}$ on 19 March 2025) and temporal (1~s) scales in both AR and QS. Owing to the lower spatial resolution of our datasets compared to the perihelion EUI $\hrieuv$ observations, the smallest detected surface areas in the QS lie within the range already reported by \citet{2025A&A...699A.138N} (approximately 0.01~$\text{Mm}^2$). Enabled by the unprecedented 1~s cadence, our observations detected very short-lived events, with lifetimes shorter than 3~s. To verify that the detection of short-lived events is genuinely due to the high temporal cadence, we rebinned the QS dataset from 19 October 2024 to lower cadences and applied the same detection method and threshold. For the dataset rebinned to 3~s, the minimum detectable lifetime was found to be 3~s, confirming that the shorter-lifetime events identified in the original data are only detectable thanks to the superior 1~s cadence (see Appendix \ref{appendix:influence_cadence} for further details). 

The shape of the lifetime and surface area distributions appears qualitatively similar between QS and AR, and the mean values of lifetime and surface area do not show substantial differences between the two regions. A similar pattern has been observed in so-called blinkers, impulsive brightening events detected in transition region lines using spectrometers, which also exhibit comparable average properties across different solar regions \citep{2002SoPh..206..249P}. Moreover, although the magnetic origins of EUV brightenings detected with EUI span a range of environments including strong bipolar, unipolar, and weak field regions, \citet{2024A&A...692A.236N} reported that the distributions of all brightenings were nearly identical to those associated with strong bipolar regions. This is consistent with our finding that, despite the expected differences in magnetic activity between AR and QS, the distributions of EUV brightenings in the two regions appear remarkably similar. This may suggest that differences in photospheric magnetic activity do not necessarily translate into significant variations in coronal dynamics. In the case of peak brightness, the AR distribution on 19 October 2024 resembles a right-skewed log-normal shape, whereas the QS distribution is closer to a symmetric log-normal form with a lower central value. On 19 March 2025, both AR and QS distributions exhibit approximately log-normal behaviour, but with distinct centres: the QS distribution peaks at lower brightness values, while the AR distribution is shifted toward higher brightness values. Overall, the peak brightness values in ARs tend to be higher than those in the QS, which can be naturally attributed to the presence of intrinsically brighter structures in ARs due to their stronger magnetic fields. The smaller separation between the AR and QS distributions on 19 October 2024 could be influenced by projection effects, as this dataset was taken closer to the limb where overlying dark structures can partially obscure lower layers, leading to an overall dimmer appearance.



\subsubsection{Birthrates of EUV brightenings}\label{sec:birthrate}

For further analysis, and following the criteria adopted in \citet{2025A&A...699A.138N}, we exclude events that occupy only a single pixel during just one timeframe. After this filtering, the number of detected events is 111~240 in the AR on 19 October 2024, 23~939 in the QS on 19 October 2024, 410~927 in the AR on 19 March 2025, and 295~725 in the QS on 19 March 2025.
By dividing these numbers by the respective observational areas and durations, we derive birthrates of $5.7\times10^{-15}\,\text{m}^{-2}\,\text{s}^{-1}$ for the AR and $2.0\times10^{-15}\,\text{m}^{-2}\,\text{s}^{-1}$ for the QS on 19 October 2024, and $1.1\times10^{-14}\,\text{m}^{-2}\,\text{s}^{-1}$ for the AR and $4.3\times10^{-15}\,\text{m}^{-2}\,\text{s}^{-1}$ for the QS on 19 March 2025. These results clearly show that, on both observation dates, the birthrates of EUV brightenings in ARs are around three times higher than those in QS regions. A similar trend has previously been reported for EUV brightenings observed with SOHO/EIT \citep{1998A&A...336.1039B, 1999SoPh..186..207B} as well as for blinkers \citep{2002SoPh..206...21B, 2002SoPh..206..249P}. We also note that the birthrates derived from the 19 October 2024 datasets, in both AR and QS, are lower by roughly a factor of two compared to those from 19 March 2025. This discrepancy may plausibly be attributed to projection effects. The 19 October 2024 observations were taken closer to the limb, where brightening events are more likely to be partially obscured by overlying coronal structures. Such geometric effects could reduce the number of detected events, leading to an underestimation of the true birthrate. Other factors may also play a role, including the slightly higher spatial resolution of the 19 March 2025 dataset and possible differences in the noise levels across datasets. However, the extent to which each of these factors influences the derived birthrates cannot be disentangled within the scope of the present study. A more definitive assessment would require controlled experiments in which these variables are isolated and systematically compared.

For comparison, QS EUV brightenings detected in EUI $\hrieuv$ data with the same spatial resolution as our 19 October 2024 dataset, but observed at a cadence of 3~s, yield a birthrate of $5.6\times10^{-17}\,\text{m}^{-2}\,\text{s}^{-1}$ \citep{2024A&A...692A.236N}. This highlights, in agreement with the histogram analysis, that a higher temporal cadence allows for the detection of significantly more events. In \citet{2025A&A...699A.138N}, QS EUV brightening birthrates of $6.3\times10^{-16}\,\text{m}^{-2}\,\text{s}^{-1}$ and $5.9\times10^{-16}\,\text{m}^{-2}\,\text{s}^{-1}$ were reported using two EUI datasets obtained at perihelion with a 3~s cadence. To directly compare with this study, we redetected events in our two QS datasets using the same detection threshold ($n=6$) as employed in \citet{2025A&A...699A.138N}. The resulting birthrates were $7.8\times10^{-16}\,\text{m}^{-2}\,\text{s}^{-1}$ on 19 October 2024 and $6.7\times10^{-16}\,\text{m}^{-2}\,\text{s}^{-1}$ on 19 March 2025. Despite the lower spatial resolution of our data, these values confirm that higher temporal resolution facilitates the detection of a larger number of events. To further assess whether this increase is solely due to the higher cadence, we recomputed the birthrates by restricting the analysis to events with lifetimes longer than or equal to 3~s. In both datasets, this yielded a consistent birthrate of $1.8\times10^{-16}\,\text{m}^{-2}\,\text{s}^{-1}$. This confirms that, when applying the same detection threshold and temporal resolution, the lower spatial resolution of our data still results in reduced birthrates. Consequently, we reaffirm that the higher temporal resolution enables the detection of numerous short-lived events that would otherwise remain undetected due to the instrumental limitations.

\subsubsection{Statistical relationships between EUV brightening characteristics}
\begin{figure}
  \resizebox{\hsize}{!}{\includegraphics{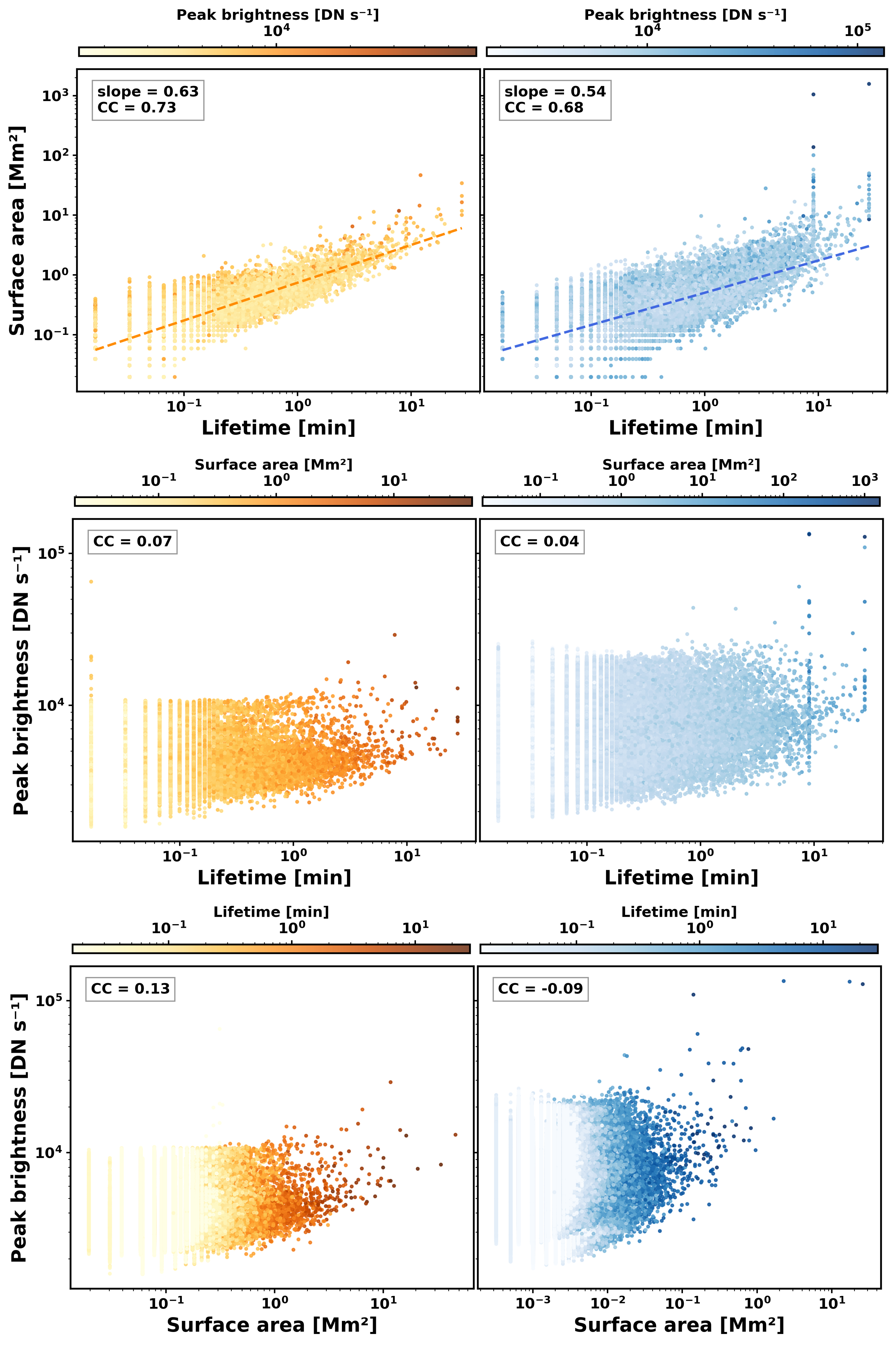}}
  \caption{Scatter plots showing the relationships between the lifetimes, surface areas, and peak brightnesses of EUV brightenings, detected in the quiet Sun (left panels) and active regions (right panels). The dashed lines represent linear fits on the log-log scale. The correlation coefficient and the slope of each linear fit are indicated in the legend of each panel.}
  \label{fig:scatter_campfires}
\end{figure}

Fig.~\ref{fig:scatter_campfires} shows scatter plots between the lifetime, surface area, and peak brightness of EUV brightenings detected in each AR and QS. The surface area and lifetime exhibit a clear linear relationship in both regions. In the QS, the slope is approximately 0.63 with a Pearson correlation coefficient (CC) of 0.73, while in the AR, the slope is 0.54 with a CC of 0.68. For comparison, EUV brightenings observed in the corona with SOHO/EIT showed a slope of 1.1 in the QS \citep{1998A&A...336.1039B} and 0.9 \citep{1999SoPh..186..207B} in the AR, showing a similar trend as ours. 

On the contrary, peak brightness shows no significant correlation with either lifetime or surface area in our datasets. Previous studies have found that around 50\% of EUV brightenings observed with EUI occur in strong bipolar photospheric magnetic regions, suggesting a potential connection to Ellerman bombs (EBs; \citealt{2018SSRv..214..120Y}), impulsive events that originate in the photosphere. EBs are known to be magnetically linked to events occurring higher in the atmosphere, such as in the transition region \citep{2019ApJ...875L..30C, 2025A&A...693A.221B}. The statistical properties of EBs observed in ARs reveal a strong correlation between peak intensity and area (CC = 0.99), and moderate correlations (CC = 0.60) between both area and lifetime, and intensity and lifetime \citep{2002ApJ...575..506G}. On the other hand, EBs observed in the QS show no clear trends among area, lifetime, and peak intensity. Rather, they appear more scattered, as seen in Figure 6 of \citet{2022A&A...664A..72J}. These comparisons indicate that the EUV brightenings identified in both AR and QS in our study do not fully match the statistical behavior of known EBs, suggesting that they may involve different physical mechanisms or formation conditions.

In the AR results, we also note the presence of vertical trends in the scatter plots, particularly at lifetimes of approximately 10 and 30~minutes. These correspond to two prominent bins in the lifetime histograms (Fig.~\ref{fig:hist_campfires}), which represent the longest lifetime bin in each dataset, respectively. Although one might attribute these features to boundary effects of the detection method, the QS datasets, analysed with the same procedure, show only a weak indication of such behaviour and not as prominently as in the ARs. A more definitive assessment would therefore require a broader investigation using additional datasets. Nevertheless, we do not expect these outliers to affect the statistical results presented in this study, and we have therefore retained them in our analysis.

\begin{figure}
  \resizebox{\hsize}{!}{\includegraphics{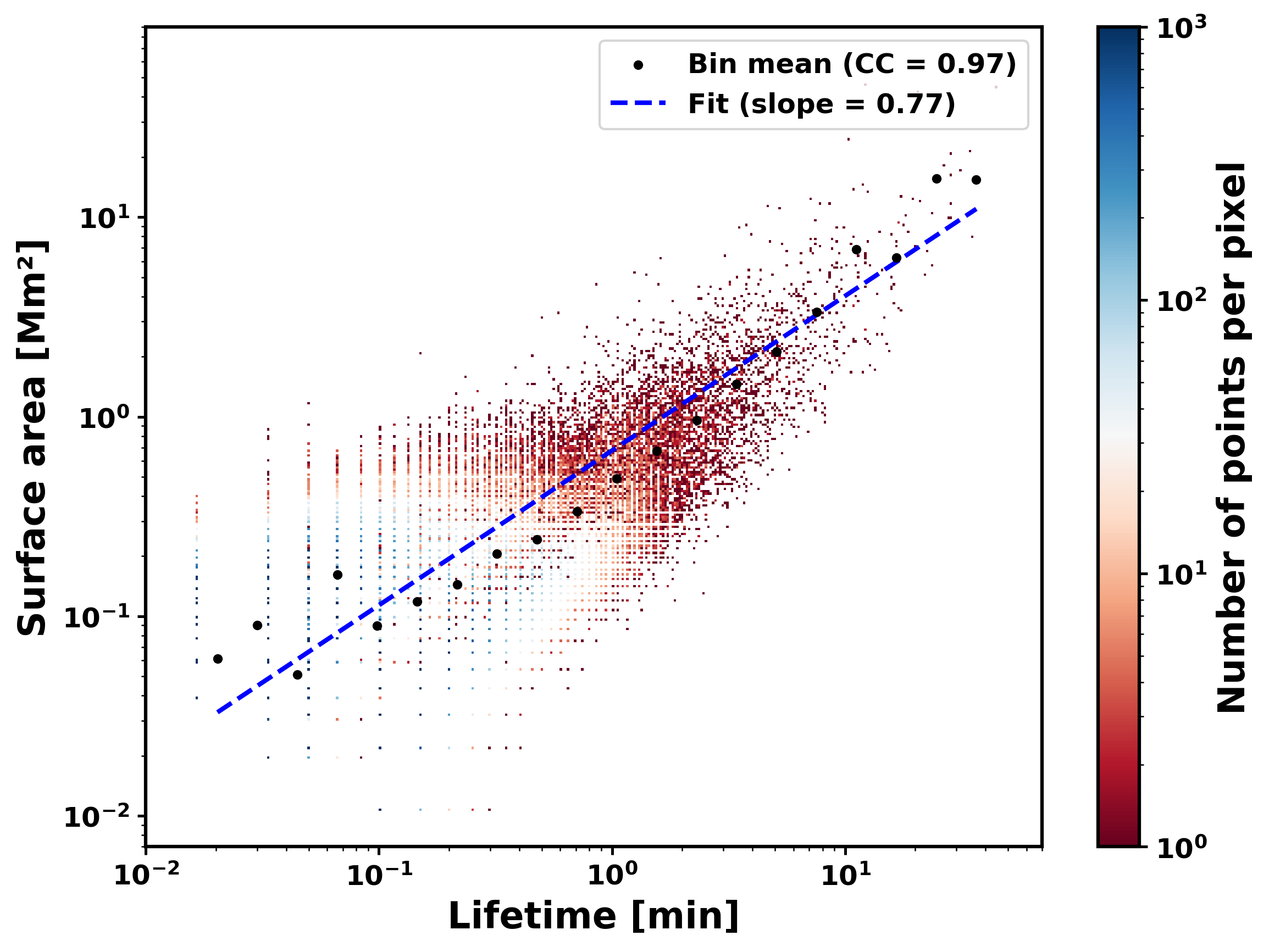}}
  \caption{Scatter density maps showing the relationship between the lifetimes and surface areas of quiet Sun EUV brightenings detected in this study, combined with the events reported in \citet{2025A&A...699A.138N}. The colour scale indicates the number density of events per pixel in logarithmic normalisation. Black points correspond to the mean surface area within each of 20 equally spaced bins in the lifetime. The blue dashed line indicates a linear fit to the binned (black) data. The correlation coefficient for the full dataset is 0.31. The correlation coefficient and slope of the linear fit to the binned data, are shown in the legend.}
  \label{fig:scatter_campfires_all}
\end{figure}

Fig.~\ref{fig:scatter_campfires_all} presents a combined analysis of the relationship between lifetime and surface area for EUV brightenings detected with the highest spatial resolution currently available from EUI observations \citep{2025A&A...699A.138N} and those detected in this study using the highest temporal resolution. To assess the presence of a linear trend in the scattered distribution, we divided the events by lifetime into 20 bins and calculated the mean surface area in each bin. This binned analysis yields a Pearson correlation coefficient of 0.97 and a slope of 0.77, indicating a robust and consistent linear trend. While our study enables access to events with very short lifetimes (towards the left of the plot), and \citet{2025A&A...699A.138N} reveals events with very small areas (towards the bottom), the lower-left region representing events with both short lifetimes and small areas remains largely unexplored. This highlights a still-inaccessible regime in the current parameter space of EUV brightenings. Future high-resolution EUI campaigns that combine both high spatial and temporal resolution are expected to probe this domain, offering deeper insights into the nature of the smallest and most transient EUV brightenings.

\section{QPPs in EUV brightenings}\label{sec:qpp}

For the QPP analysis, we selected 64~526 AR brightenings and 28~354 QS brightenings with lifetimes equal to or greater than five time frames (5~s), which constitutes a stricter threshold than the Nyquist criterion of two time frames. After applying the QPP detection methods, we estimated the number of oscillation cycles by dividing the signal lifetime by its period, and excluded cases with fewer than two cycles from the data set. The detected EUV brightenings exhibit dynamic evolution in both spatial extent and position throughout their lifetimes. To investigate potential QPP signatures, we constructed normalised light curves by summing the brightness within the region corresponding to the maximum spatial extent of each event. The integration was performed over the lifetime of the event in order to accurately capture the full temporal evolution. 

\begin{figure*}
  \resizebox{\hsize}{!}{\includegraphics{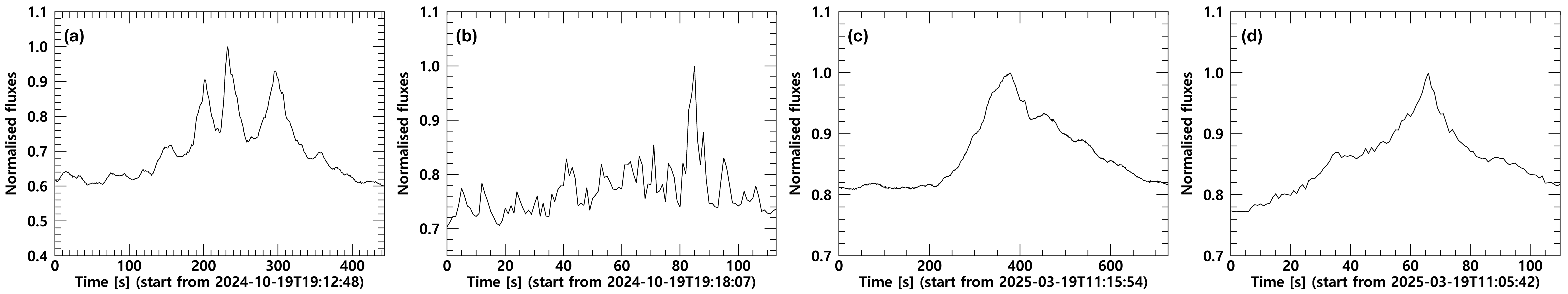}}
  \caption{Integrated light curves of representative EUV brightenings shown in Fig.~\ref{fig:campfires}, with each panel corresponding to one of the events.}
  \label{fig:lightcurves}
\end{figure*}

Representative examples of the resulting integrated light curves are shown in Fig.~\ref{fig:lightcurves}, demonstrating notable variations in shape depending on the size and lifetime of the event. Among the four examples, one event exhibits a single significant peak (Fig.~\ref{fig:lightcurves} c), while another displays multiple pronounced peaks (Fig.~\ref{fig:lightcurves} a). This diversity in light curve morphology is reminiscent of the intensity profiles observed in blinkers, which also show similar classifications \citep{2004A&A...422..709B}. Visual inspection of the entire sample revealed no systematic differences in light curve morphology between brightenings occurring in ARs and those in QS regions.

\subsection{Analysis}

QPPs feature a wide range of behaviours, from highly periodic and stationary signals \citep{2020ApJ...895...50H} to more complex, quasi-periodic, and non-stationary patterns \citep{2019PPCF...61a4024N}. In this study, to maximise the detection of QPPs regardless of their characteristics, we adopted two analysis techniques, each optimised for different types of oscillatory behaviour \citep{2019ApJS..244...44B}.

\subsubsection{Stationary QPPs}
The Fourier analysis-based detection method known as Automated Flare Inference of Oscillations\footnote{\url{https://github.com/aringlis/afino_release_version}} (AFINO; \citealt{2015ApJ...798..108I, 2016ApJ...833..284I}) is specifically designed to identify stationary types of QPPs that are nearly periodic in nature. It has been successfully employed to detect QPPs occurring in EUV brightenings \citep{2025A&A...698A..65L}, solar flares \citep{2020ApJ...895...50H}, and stellar flares \citep{2025A&A...700A.178J}, and has also been demonstrated to be applicable to the detection of very short-period QPPs \citep{2024ApJ...971...29I}. One of AFINO’s main advantages is that it eliminates the need for preprocessing, such as detrending, which can produce misleading results \citep{2016ApJ...825..110A, 2018SoPh..293...61D}. It is also well suited to efficiently process large datasets in a statistically rigorous way. A comprehensive description of AFINO is provided in \citet{2015ApJ...798..108I, 2016ApJ...833..284I} and \citet{2024ApJ...971...29I}. Below, we briefly summarise the key aspects of the method as relevant to the present study.

\begin{figure*}
  \resizebox{\hsize}{!}{\includegraphics{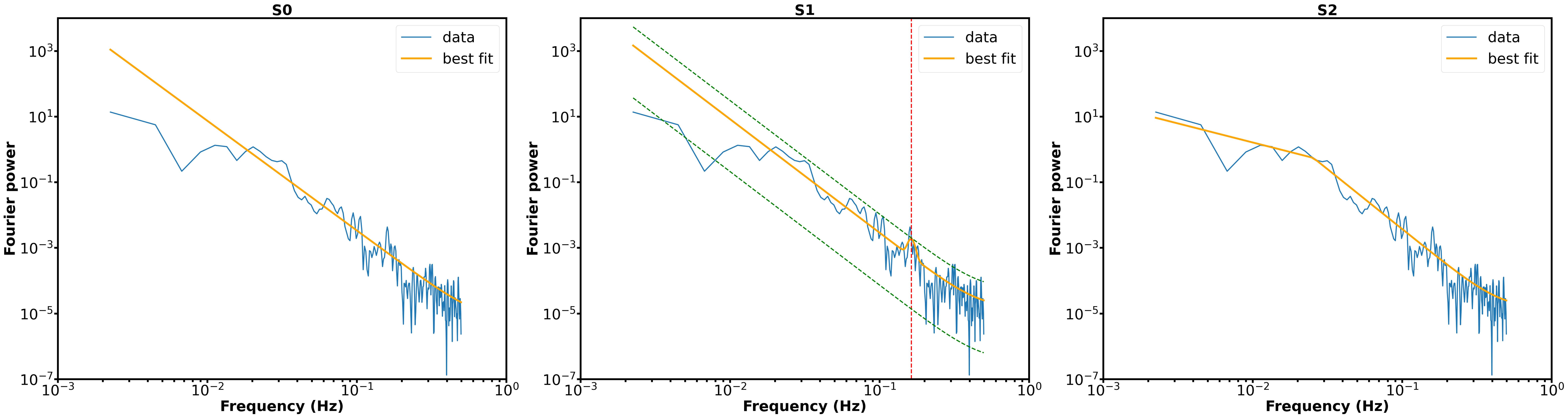}}
  \caption{An example of AFINO results applied to the integrated light curve of the EUV brightening shown in Fig.~\ref{fig:lightcurves} a. The blue line represents the Fourier power of the light curve, while the orange line indicates the best fit for each model: $S_{0}$ (left), $S_{1}$ (centre) and $S_{2}$ (right). The red vertical dashed line in Model $S_{1}$ marks the Gaussian centre corresponding to the dominant period, and the green dashed lines represent the 95\% confidence interval. The Bayesian Information Criterion (BIC) differences are $\Delta\text{BIC}_{S_{0}-S_{1}} = -1.1$ and $\Delta\text{BIC}_{S_{2}-S_{1}} = -15.3$. Model $S_{2}$ provides the best fit to the Fourier power.}
  \label{fig:afino}
\end{figure*}

AFINO identifies QPP signatures by fitting models to the Fourier power spectrum of flare light curves. Specifically, it compares three competing models: a simple power law ($S_{0}$), a power law with a Gaussian bump representing a strong periodicity ($S_{1}$), and a broken power law ($S_{2}$). The selection of the best-fitting model is guided by the Bayesian information criterion (BIC), which evaluates the trade-off between model complexity and goodness of fit. Among the three models considered, Model $S_{1}$ is favoured when it yields lower BIC values than both S0 and S2, indicating that the light curve contains a significant periodic component. Specifically, we consider a QPP candidate valid when 
$\Delta\text{BIC}_{S_{0}-S_{1}}>0$ and $\Delta\text{BIC}_{S_{2}-S_{1}}>0$, suggesting that the periodic model ($S_{1}$) provides a statistically superior representation of the data. To ensure the reliability of the fit, we additionally require that Model S1 has a sufficiently high reduced $\chi^2$ value (greater than 0.01). Fig.~\ref{fig:afino} presents an example of the AFINO result for one of the EUV brightenings shown in Fig.~\ref{fig:lightcurves} a. The BIC differences were calculated as $\Delta\text{BIC}_{S_{0}-S_{1}} =-1.1$ and $\Delta\text{BIC}_{S_{2}-S_{1}}=-15.3$. Since the BIC for Model $S_{1}$ was lower than the two other models, this EUV brightening was not classified as exhibiting a stationary QPP event.

\subsubsection{Non-stationary QPPs}
Ensemble empirical mode decomposition (EEMD; \citealt{wu09}) has been shown to be well suited for the detection and analysis of quasi-periodic and non-stationary oscillatory signatures. It has been successfully employed to identify non-stationary QPPs in both solar and stellar flares \citep{2015A&A...574A..53K, 2016ApJ...830..110C, 2022SSRv..218....9A}. EEMD decomposes a signal into a set of intrinsic mode functions (IMFs). Accordingly, in this study, we employed EEMD\footnote{\url{https://pyemd.readthedocs.io/en/latest/examples.html}} to analyse the light curves of EUV brightenings by separating them into their constituent IMFs. 

\begin{figure*}
  \resizebox{\hsize}{!}{\includegraphics{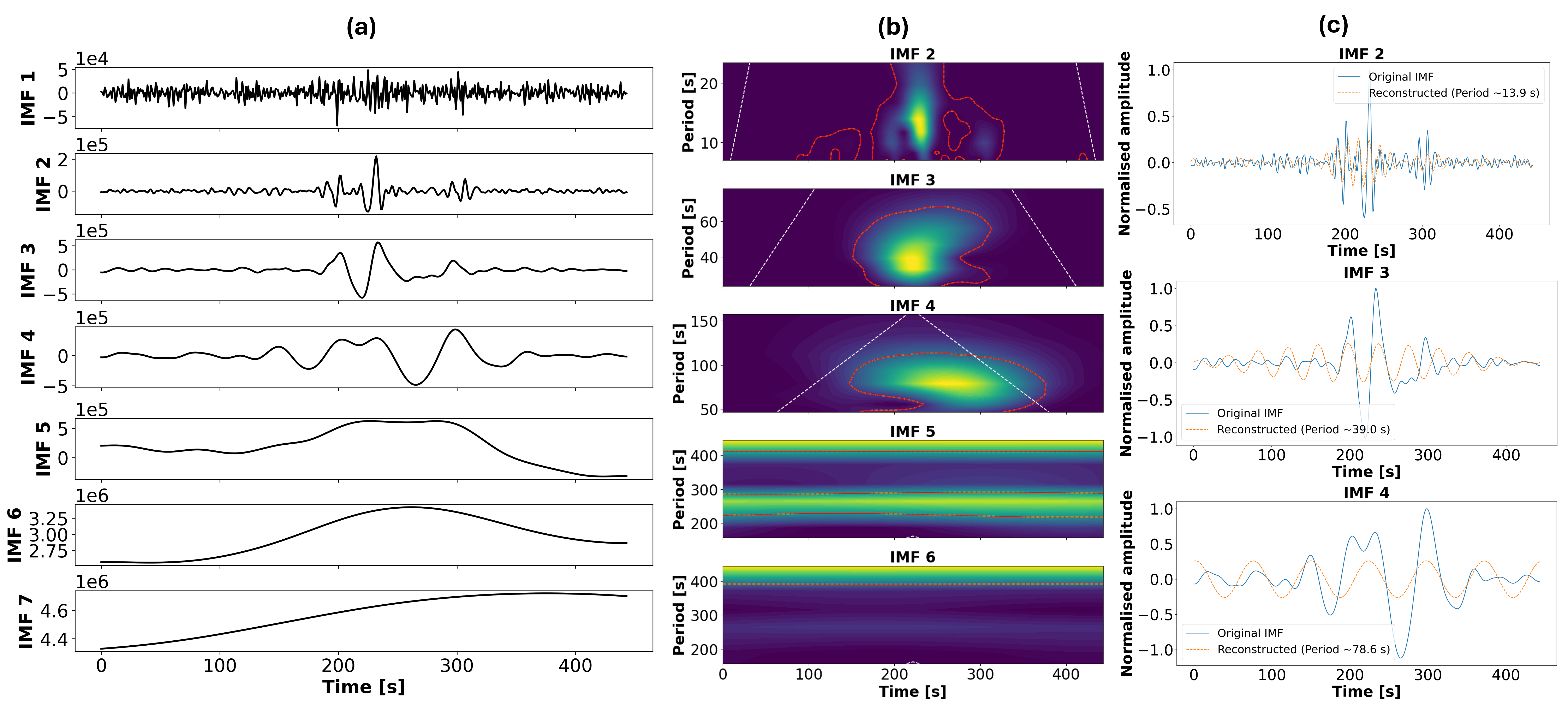}}
  \caption{An example of ensemble empirical mode decomposition (EEMD) and wavelet analysis applied to the integrated light curve of the EUV brightening shown in Fig.~\ref{fig:lightcurves}a. (a) Intrinsic mode functions (IMFs) obtained using the EEMD technique. (b) Wavelet power spectra for the 2nd to 6th IMFs. Darker to lighter colours indicate increasing power. The red solid line denotes the 95\% significance level based on a red noise background, while the yellow dashed line corresponds to the 95\% significance level for white noise. The white dashed curve outlines the cone of influence. (c) Comparison between the original IMF signals (blue) and their corresponding narrowband signals (orange dashed) reconstructed from the Fourier power spectrum at the dominant periods of 13.9, 39, and 78.6 seconds, identified in the 2nd to 4th IMFs.}
  \label{fig:eemd}
\end{figure*}

The left panel of Fig.~\ref{fig:eemd} shows an example of EEMD results applied to a brightening in which no stationary QPP was identified using the AFINO method (see Fig.~\ref{fig:afino}). The original light curve was decomposed into a total of seven IMFs. Following previous studies that utilised the IMF with the slower characteristic timescale to represent the long-term trend of the original signal \citep{2015A&A...574A..53K, 2016ApJ...830..110C}, we adopted the same approach. For this particular example, we could manually determine that a combination of the 5th to 7th IMFs captured the trend of the light curve, while the first IMF was likely to correspond to high-frequency noise. Although manual classification allows accurate identification of trend and noise components for individual cases, it is clearly infeasible to apply such an approach to the full dataset comprising 64~526 AR and 28~354 QS brightenings. Therefore, for the bulk analysis, we assumed the first IMF to represent noise and the final IMF to represent the trend. The remaining IMFs were then subjected to wavelet analysis, with the aim of identifying statistically significant periodicities.

The centre panel of Fig.~\ref{fig:eemd} displays the wavelet power spectra computed from the IMFs, excluding the first and last. When taking into account both red and white noise backgrounds, only the second, third, and fourth IMFs displayed power exceeding the 95\% confidence level and lying outside the cone of influence. The periods identified in these IMFs were 13.9, 39, and 78.6 seconds, respectively. This result also confirms that the fifth and sixth IMFs, which likely form part of the long-term background trend, were appropriately excluded from further consideration. To validate whether the identified periodicities were genuinely representative of the respective IMF signals, we compared each IMF with the corresponding narrowband signal reconstructed from the Fourier power spectrum. As illustrated in the right panel of Fig.~\ref{fig:eemd}, the dominant periods revealed in the wavelet analysis were in good agreement with the temporal evolution of each IMF over the relevant intervals. Through this multi-step approach, we concluded that the examined EUV brightening exhibits multi-mode QPPs with intrinsic periods of 13.9, 39, and 78.6~s, consistent with previous findings of multi-periodic QPPs in solar flares \citep{2015A&A...574A..53K}. This same methodology was systematically applied to the entire dataset. 

We note that accurately capturing non-stationary (i.e., time-varying) signals ideally requires tracking how detected periodicities evolve across different phases of the EUV brightening light curve \citep{2023MNRAS.523.3689M}. In our wavelet results, the intervals of significant power do not always coincide perfectly across IMFs, implying potential temporal variation in periodicity. However, we refrain from explicitly classifying such cases as time-varying signals based on this criterion alone. Unlike solar or stellar flares, EUV brightenings still lack clearly defined phases, and their temporal profiles remain less well characterised. This section aims to supplement the AFINO-based analysis by detecting significant periodicities that are localised in time and therefore may be overlooked by methods assuming stationarity over the entire event duration. To achieve this, we adopt a widely used time-frequency approach and, for simplicity, classify events with significant but temporally localised periodicities (detected via wavelet or EEMD) as non-stationary QPPs. Future studies incorporating phase-resolved analysis of EUV brightening light curves would help further clarify the nature of these time-localised QPP signatures and distinguish between different underlying physical mechanisms.

\subsection{Results and discussion}

Following the application of the detection methods outlined above to the light curves of 64~526 AR and 28~354 QS brightenings, we identified a total of 477 stationary QPPs and 6~313 non-stationary QPPs within ARs. Among these, 262 EUV brightening events exhibited both stationary and non-stationary QPP signatures. This proportion is comparable to previous findings based on Geostationary Operational Environmental Satellite (GOES) X-ray observations, where approximately 48\% of 205 M- and X-class flares that exhibited stationary QPPs also showed non-stationary components \citep{2023MNRAS.523.3689M}. In QS regions, 178 stationary QPPs and 2~302 non-stationary QPPs were detected. A total of 78 EUV brightenings were found to contain both types of QPPs, representing a slightly lower proportion compared to the AR cases.


\subsubsection{Dependence of QPP occurrence on EUV brightening properties}

To investigate how the physical properties of EUV brightenings influence the occurrence of QPPs, we adopted the binning methodology introduced by \citet{2025A&A...698A..65L}, in which the dynamic range of each parameter was divided into five logarithmically spaced groups. Given that the range of parameters in the present dataset differs from that examined in the earlier study, we redefined the bin boundaries accordingly for the AR and QS regions and calculated the average occurrence rates based on the updated distributions. These bins were designated as follows: for surface area, Group A (smallest) to Group E (largest); for lifetime, Group~$\alpha$ to Group~$\epsilon$; and for peak brightness, Group I to Group V. While the group labels remain consistent with those previously defined, their corresponding value intervals have been updated to reflect the present dataset.

\begin{figure*}
  \resizebox{\hsize}{!}{\includegraphics{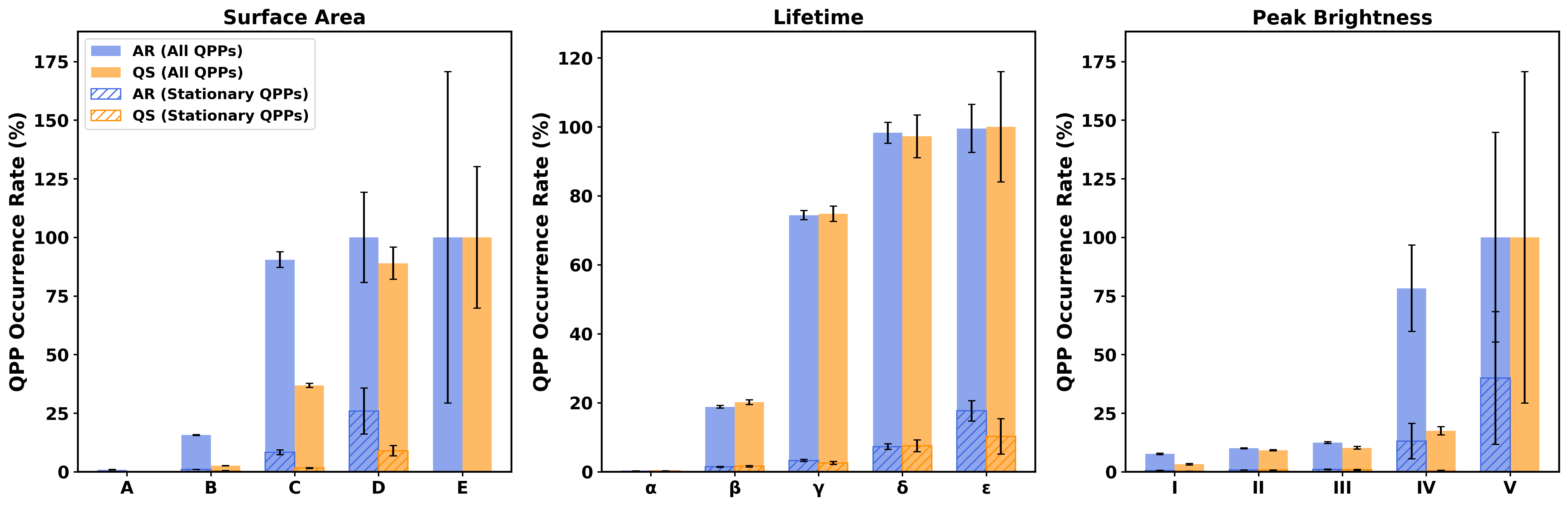}}
  \caption{QPP occurrence rates as a function of the properties of EUV brightenings: surface area (left), lifetime (middle), and peak brightness (right). Blue and orange bars represent brightenings in ARs and QS regions, respectively. Coloured bars show the total occurrence rates of QPPs (both stationary and non-stationary), while hatched bars indicate stationary QPPs only. The error bars, which account for the varying bin populations, were calculated based on Poisson statistics.}
  \label{fig:Rates}
\end{figure*}

As shown in Fig.~\ref{fig:Rates}, the QPP occurrence rate generally increases with surface area, lifetime, and peak brightness in both AR and QS regions. For each parameter, the two largest bins exhibit nearly 100\% occurrence. The occurrence rates of stationary QPPs represent only a small fraction of the total, and their distribution across parameter groups is consistent with those found in the QS EUV brightenings observed at a 3~s cadence using the same AFINO method \citep{2025A&A...698A..65L}. Across all parameter bins, the differences in occurrence rates between AR and QS are minimal, though a slightly higher tendency is observed in the ARs.

\begin{figure*}
  \resizebox{\hsize}{!}{\includegraphics{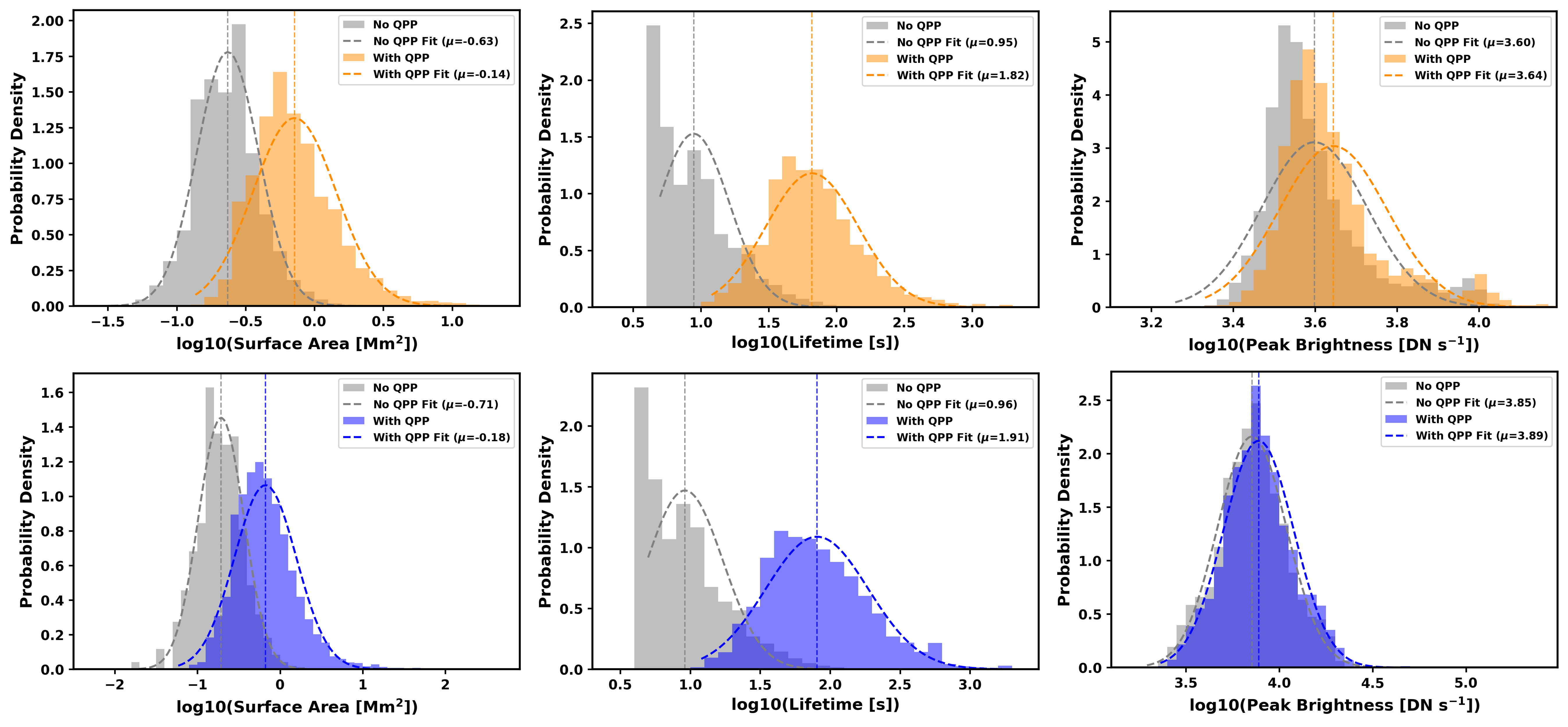}}
  \caption{Probability density distributions of three parameters, surface area (left), lifetime (middle), and peak brightness (right), for EUV brightenings with (orange or blue) and without (grey) detected QPPs. The top row corresponds to QS brightenings and the bottom row to AR brightenings. The distributions are shown in log-scale for each parameter. Overplotted dashed curves represent Gaussian fits to each distribution, with vertical dashed lines indicating the mean values ($\mu$) from the fits.}
  \label{fig:hist_QPPType}
\end{figure*}

To further quantify the relative influence of the three parameters, we performed a multivariate logistic regression using surface area, lifetime, and peak brightness as explanatory variables. While the bin-wise occurrence rates suggest that all three parameters may influence QPP occurrence, they are not mutually independent. The regression results demonstrate that lifetime is the most significant factor, with a strong and statistically significant positive coefficient, indicating that brightenings with longer durations are substantially more likely to show QPPs. On the contrary, surface area shows a weak but statistically significant negative correlation, and peak brightness exhibits no statistically significant effect. These findings are further supported by comparing the mean values of each parameter between brightenings with and without detected QPPs (see Fig.~\ref{fig:hist_QPPType}). Notably, the difference in mean lifetime is the most pronounced, whereas the mean peak brightness shows virtually no difference between the two groups.

However, it is important to recognise that the observed lifetime dependence may be influenced, at least in part, by detection-related biases. Both Fourier- and wavelet-based detection methods require a minimum signal duration to reliably identify periodic signatures. As a result, brightenings with shorter lifetimes may in fact exhibit QPPs, but their brevity may prevent those periodicities from being robustly detected. Consequently, the strong lifetime dependence revealed by the regression analysis may reflect a methodological limitation rather than a true physical preference.

\subsubsection{The periods of QPPs}
The detected periods of stationary QPPs range from approximately 5 to 300~s in both AR and QS regions. For non-stationary QPPs, the periods extend more broadly from 5 to about 530~s in ARs, and from 5 to about 400~s in QS. The minimum detectable period of 5~s arises from the selection criterion requiring a minimum of five time frames (with 1~s cadence) to ensure reliable identification of periodic signals. For reference, the minimum QPP period detectable in previous studies using a 3~s cadence was approximately 15~s \citep{2025A&A...698A..65L}. These results collectively suggest that EUV brightenings likely host short-period QPPs down to the instrument's temporal resolution, many of which may remain undetectable in lower-cadence observations due to sampling limitations.

\begin{figure*}
  \resizebox{\hsize}{!}{\includegraphics{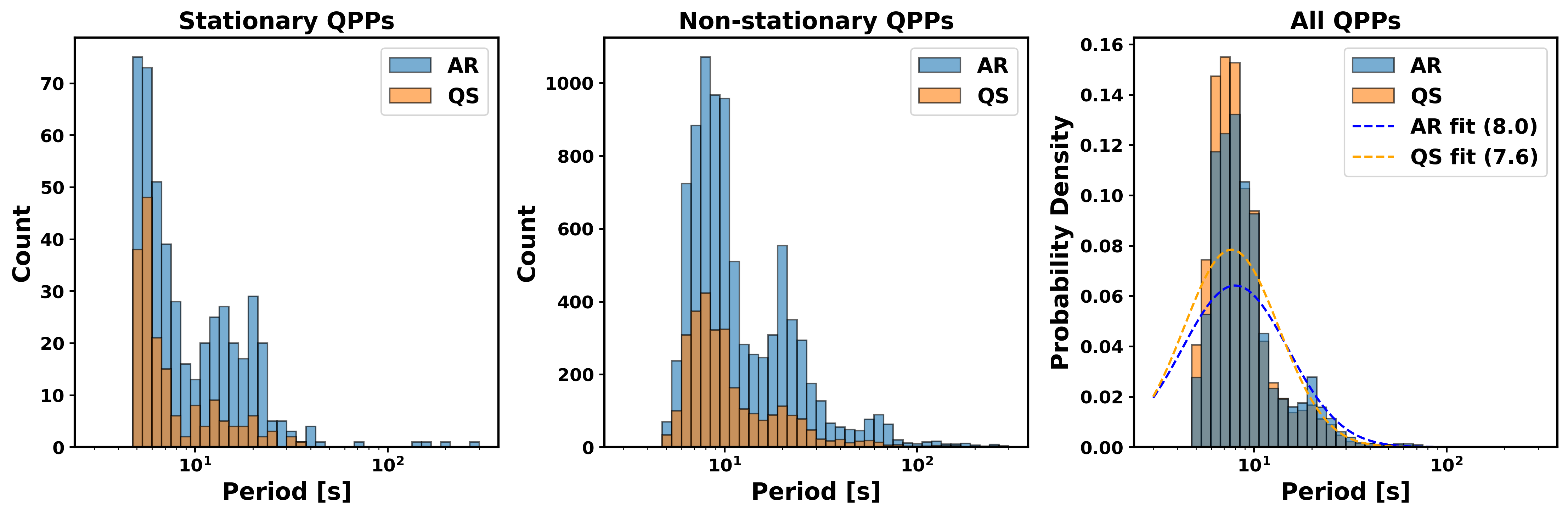}}
  \caption{Histograms of QPP periods detected in EUV brightenings. The left and middle panels show the distributions for stationary and non-stationary QPPs in ARs (blue) and QS regions (orange), respectively. The right panel displays the combined period distributions with log-normal fits, and the mean values are indicated in the legend.}
  \label{fig:hist_per}
\end{figure*}

While the overall period distributions are broadly similar between stationary and non-stationary QPPs, the former (left panel of Fig.~\ref{fig:hist_per}) tend to favour shorter periods. Interestingly, both distributions exhibit a secondary peak near 20~s. A similar enhancement has also been reported in the period distribution of QPPs observed in GOES X-ray flares \citep{2020ApJ...895...50H}, implying the existence of a preferred timescale that spans a wide range of energy regimes from small-scale EUV brightenings to large-scale flares. This cross-scale coherence may reflect the presence of scale-invariant physical processes, possibly linked to turbulent dynamics such as the Kelvin–Helmholtz instability in flaring loops \citep{2018A&A...618A.135R, 2019ApJ...877L..11R}. However, the bimodal appearance could also partly arise from the detection method itself (e.g., the 1~s cadence and minimum 5 time frame requirement).

The QPP period distributions in ARs and QS show striking similarity. When fitted with log-normal functions, following the approach of \citet{2020ApJ...895...50H} and \citet{2025A&A...698A..65L}, the mean values of the fits are approximately 8.0~s and 7.7~s, respectively. However, the overall shape of the distributions appears to follow a power-law convolved with a cut-off imposed by the observational resolution limit. Considering the substantial differences in magnetic field strength and background temperature between AR and QS regions, this resemblance suggests that the dominant QPP periods are not strongly influenced by these parameters. Numerical simulations have shown that periods produced by oscillatory reconnection decrease with increasing coronal magnetic field strength following a power-law relationship \citep{2023ApJ...943..131K}, and that their dependence on magnetic field becomes negligible when the field is sufficiently strong \citep{2024ApJ...975...10S}. Based on these results, if the similar dominant periods observed in AR and QS originate from oscillatory reconnection, it may imply that the brightenings in both regions are associated with relatively strong magnetic fields. Additionally, ubiquitous kink waves in the solar corona have been shown to exhibit similar period distributions in both AR and QS regions \citep{2024A&A...689A..16L}. These findings support the possibility that both MHD wave modes and oscillatory reconnection may be responsible for generating QPPs in EUV brightenings. 

\subsubsection{Statistical relationships between QPP periods and EUV brightening characteristics}
\begin{figure*}
  \resizebox{\hsize}{!}{\includegraphics{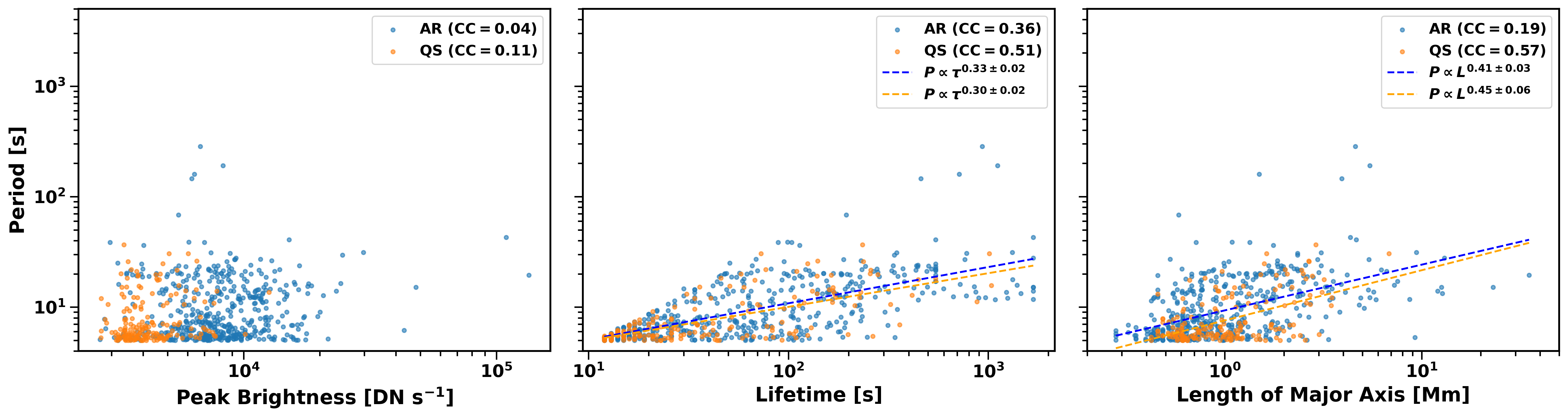}}
  \resizebox{\hsize}{!}{\includegraphics{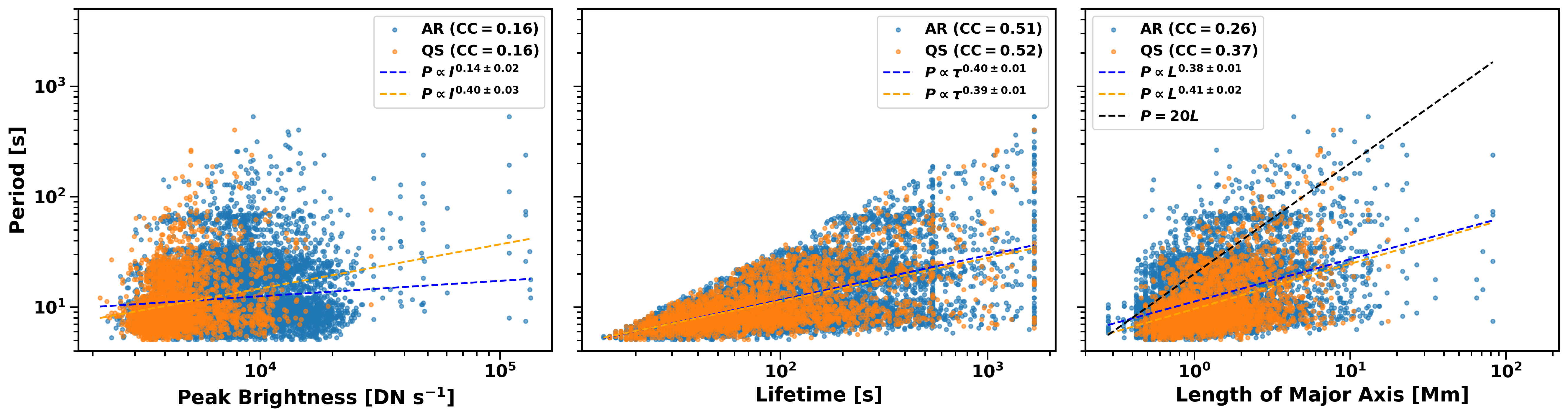}}
  \caption{Scatter plots showing the relationships between the QPP period ($P$) and the peak brightness ($I$), lifetime ($\tau$), and major axis length ($L$) of EUV brightenings in AR (blue) and QS (orange) regions. The top panels correspond to stationary QPPs, while the bottom panels show non-stationary QPPs. Pearson correlation coefficients (CC) for AR and QS are indicated in each panel. For cases where the linear regression analysis yields a statistically significant result (p-value < 0.05), the corresponding fit is shown as a dashed line, and the power-law exponent is annotated in the panel.}
  \label{fig:scatter_QPP}
\end{figure*}

Fig.~\ref{fig:scatter_QPP} presents scatter plots showing the relationships between QPP periods and three parameters of the associated EUV brightenings: peak brightness, lifetime, and the length of the major axis. Among these, the correlations between stationary QPP periods and QS event lifetime, stationary QPP periods and QS event length scale, and non-stationary QPP periods and QS and AR event lifetimes exceed a CC of 0.5, whereas all others fall below this threshold. In previous studies of stationary QPPs in QS EUV brightenings, such correlations were weaker, with CC values below 0.1 \citep{2025A&A...698A..65L}, suggesting that the present results represent a slight improvement. This improvement may partly stem from the inclusion of shorter-period (and correspondingly shorter-lived) events identified in this study, which extend the lower bound of the period distribution beyond that considered in earlier works.

Despite the low CCs, linear regression analysis reveals statistically significant relationships (p-value < 0.05) in all cases except between stationary QPP periods and peak brightness. The derived scaling laws exhibit broadly consistent trends across AR and QS regions for both lifetime and length scale, while their relationships with peak brightness diverge more noticeably. This divergence is likely related to the fact that the distributions of lifetime, surface area, and QPP periods are similar between AR and QS, whereas the peak brightness distribution in ARs is skewed towards higher values. Furthermore, the dependence of QPP period on length scale appears similar for stationary and non-stationary QPPs, whereas the period–lifetime scaling shows greater variability between the two types. This discrepancy may suggest that the two types of QPPs are driven by different physical mechanisms or follow distinct temporal evolution patterns. However, it cannot be ruled out that this difference may also be influenced by detection-related biases, as signal duration strongly affects the sensitivity of period detection methods.

The consistently weak correlations between QPP period and peak brightness (CC < 0.16) mirror similar findings from both stellar flare observations \citep{2025A&A...700A.178J} and GOES X-ray flares \citep{2020ApJ...895...50H}, where the period showed no dependence on flare energy or peak flux. This aligns with theoretical work on oscillatory reconnection, which produces periodicities independent of the amount of the energy released \citep{2022ApJ...933..142K}, reinforcing its potential as a plausible mechanism for the QPPs detected in this study.

The period–lifetime scaling exponent for QPPs (ranging from 0.30 to 0.40) is lower than that reported for GOES flares (0.67; \citealt{2020ApJ...895...50H}), but closely matches the exponent derived from Transiting Exoplanet Survey Satellite (TESS) stellar flares (0.34; \citealt{2025A&A...700A.178J}). We would like to note that the apparent boundary extending from the lower left to the upper right is likely a methodological effect, arising because the finite signal lifetime intrinsically limits the maximum detectable period. Similarly, the period–length scale exponent obtained here (0.38–0.45) falls below the corresponding value for GOES flares (0.55; \citealt{2020ApJ...895...50H}). Notably, the distribution also shows a trend following the relation $P = 20L$ (black dashed line), which assumes a standing wave with a phase speed of 100~km $\text{s}^{-1}$. If we interpret this phase speed as the sound speed of slow waves, which can naturally explain intensity variations, the implied plasma temperature would be about 0.4~MK. This is consistent with the finding that small-scale brightenings detected by $\hrieuv$ are dominated by plasma emission at chromospheric and transition-region temperatures \citep{2024A&A...688A..77D}. Another wave mode that can readily explain QPPs is the sausage mode. Similar number densities of EUV brightenings ($\sim2\times10^{10}$~cm$^{-3}$) have been independently derived from both observational \citep{2024A&A...688A..77D} and numerical \citep{2021A&A...656L...7C} studies. Based on this value, a coronal magnetic field strength of about 20~G would be required for this phase speed to correspond to the Alfv\'{e}n speed. \citet{2024A&A...690A.242M} reported on 126 coronal bright points \citep{2019LRSP...16....2M}, which may represent a phenomenon similar to EUV brightenings, and found average loop heights of about 4~Mm with magnetic field strengths around 23~G. From stereoscopic observations of EUV brightenings detected with both $\hrieuv$ and AIA, event heights were estimated to range from 1 to 5~Mm \citep{2021A&A...656A..35Z}. Assuming a similar height range for the events studied here, the inferred magnetic field strength appears consistent. While fast kink waves are generally regarded as nearly incompressible, they may still manifest as intensity fluctuations in imaging data due to line-of-sight effects \citep{2003A&A...397..765C}, or because loop displacement associated with kink waves causes the structures to move into and out of a spatial pixel, leading to intensity variations in spectral data \citep{2012ApJ...759..144T}. The coronal loop length–period relation observed for kink waves (vertical branch in Fig. 5 of \citealt{2024A&A...685A..36S}) also reveals a linear trend corresponding to a phase speed of about 100~km~s$^{-1}$. Several mechanisms have been proposed to explain this branch, including propagating waves driven by photospheric motions \citep{2023ApJ...955...73G}, signatures of slow-mode oscillations \citep{2024MNRAS.527.5741L}, or observational undersampling effects \citep{2024A&A...690L...8L}. These raise the possibility that at least the events following this particular scaling relation may be linked to MHD waves. Nevertheless, further targeted investigations are needed to confirm these interpretations.

\begin{figure}
  \resizebox{\hsize}{!}{\includegraphics{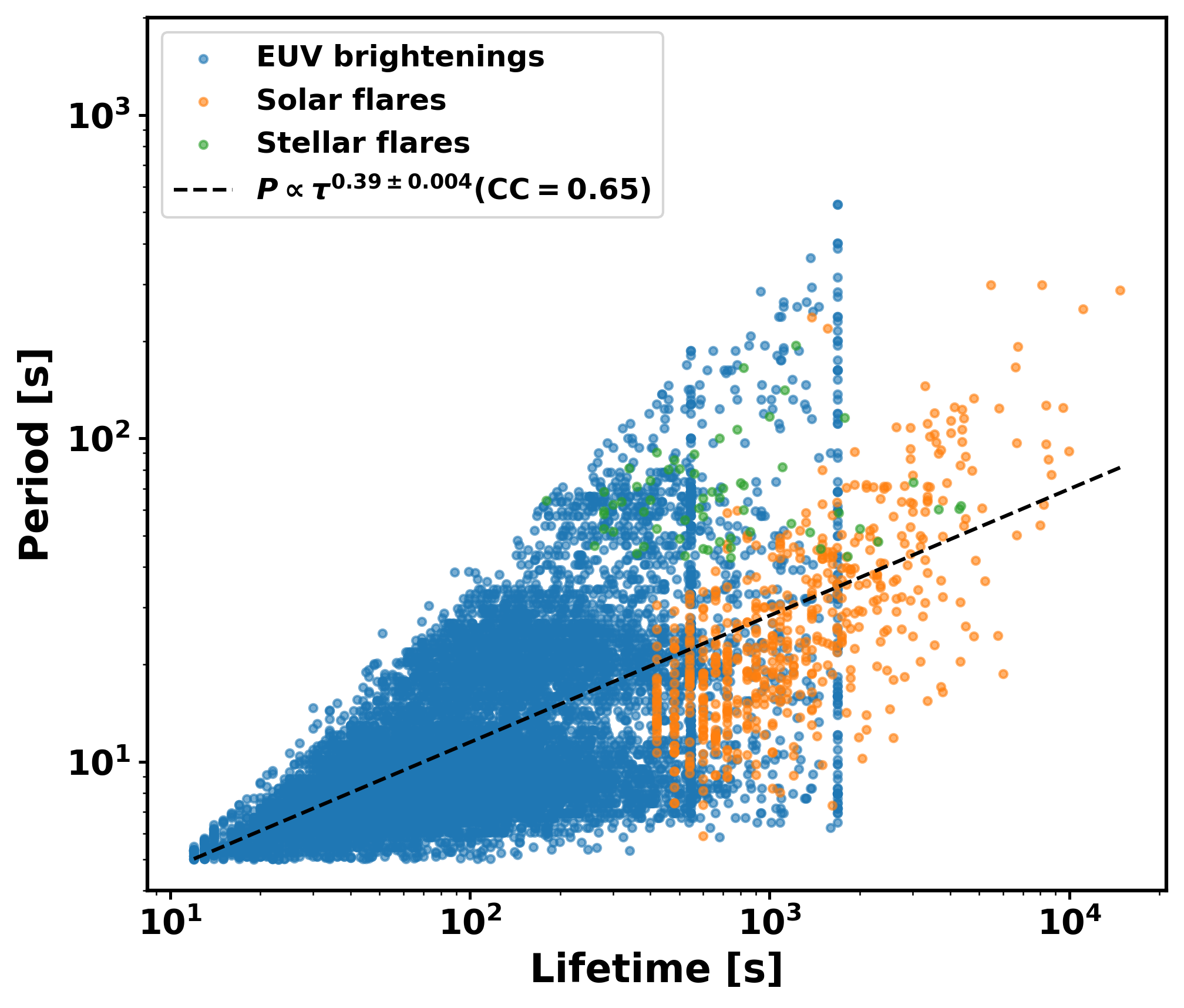}}
  \caption{Relationship between QPP period ($P$) and lifetime ($\tau$) across different flare scales. Blue, orange, and green dots represent QPPs detected in EUV brightenings (this study), GOES X-ray solar flares \citep{2020ApJ...895...50H}, and TESS stellar flares \citep{2025A&A...700A.178J}, respectively. The dashed black line shows a power-law fit to the combined dataset.}
  \label{fig:scatter_QPP_all}
\end{figure}

To examine whether a unified scaling law might apply across different flare magnitudes, we compared our results with all QPPs (both stationary and non-stationary QPPs occurring in ARs and QS regions) previously detected in GOES X-ray flares \citep{2020ApJ...895...50H} and TESS stellar flares \citep{2025A&A...700A.178J}. As shown in Fig.~\ref{fig:scatter_QPP_all}, the combined dataset exhibits a consistent trend, analogous to the well-established temperature–emission measure scaling from EUV brightenings to stellar flares \citep{2023MNRAS.522.4148K}. This further supports the interpretation that EUV brightenings may be driven by mechanisms similar to those responsible for solar and stellar flares \citep{2021A&A...656L...7C, 2022SoPh..297..141B}. The resulting period–lifetime relation, with a best-fit exponent of approximately 0.39 and a CC of 0.65, suggests that this empirical scaling persists across a wide dynamic range. This indicates that (1) QPPs may be consistent with universal scaling behaviour across flare energies, (2) the period–lifetime relation could serve as a diagnostic tool for probing flare dynamics, and (3) the observed scaling laws can offer a valuable benchmark for testing theoretical models of QPP generation.

\section{Conclusions}\label{sec:conclusion}
We analysed the properties of EUV brightenings and their QPPs with the Solar Orbiter/EUI $\hrieuv$ at an unprecedented 1~s cadence. By applying the automated detection algorithm to AR and QS datasets acquired on 19 October 2024 and 19 March 2025, we identified over 500~000 events from ARs and 300~000 events from QS regions. The detected brightenings in both AR and QS regions exhibit power-law distributions in lifetime and surface area, with steeper slopes in QS. This suggests that smaller and shorter-lived events are more frequent in QS, although the overall distribution shapes are remarkably similar across regions. Thanks to the high cadence, we were able to detect a significant population of very short-lived brightenings (lifetime < 3~s), highlighting the importance of temporal resolution for capturing fine-scale coronal dynamics. Birthrates of brightenings in ARs are consistently about three times higher than in QS.

We identified QPPs in approximately 11\% of AR brightenings and 9\% of QS brightenings using two complementary techniques designed to detect both stationary and non-stationary signals. Non-stationary QPPs were found to be more common than stationary ones, and both types appeared in both AR and QS regions with broadly similar statistical properties. The QPP periods range from 5 to over 500~s. Notably, the period distributions are similar across AR and QS, implying that neither magnetic field strength nor background plasma conditions are the dominant factors controlling the QPP timescales. Statistically significant, though weak, correlations were found between QPP period and both event lifetime and spatial scale, while no dependence was observed with peak brightness.

By combining our results with previously reported QPPs in GOES solar flares and TESS stellar flares, we found that the period–lifetime relation follows a single power-law scaling across more than three orders of magnitude. The best-fit exponent of approximately 0.39 indicates that the observed relation is consistent with a universal, scale-invariant behaviour, pointing to a shared physical origin. These findings further support the interpretation that small-scale EUV brightenings may be manifestations of flare-like processes.

Given the potential significance of EUV brightenings in the context of coronal heating, our findings demonstrate that high temporal resolution, in conjunction with high spatial resolution, is crucial for revealing the full extent of small-scale, short-duration activity in the solar corona. Our statistical results provide indications that both oscillatory reconnection and MHD wave modes may contribute to the generation of QPPs in EUV brightenings. An EUV brightening, defined here as an increase in intensity, most likely reflects changes in plasma density and/or temperature within the EUI’s response range. Yet the specific physical mechanism responsible for each event remains uncertain. Resolving this issue will require further observational efforts that combine the high spatial and temporal resolution capabilities of EUI with simultaneous multi-wavelength coverage, particularly from instruments such as the Interface Region Imaging Spectrograph \citep{2014SoPh..289.2733D} and the Atacama Large Millimeter/submillimeter Array \citep{2020A&A...635A..71W}. To fully resolve the shortest-lived brightenings and rapid QPPs, future EUV imagers capable of sub-second cadence will be essential, complementing these coordinated observations.

Forthcoming missions offering high spatio-temporal resolution, such as the Multi-slit Solar Explorer \citep{2022ApJ...926...52D}, the Solar-C EUV High-Throughput Spectroscopic Telescope \citep{2019SPIE11118E..07S}, and the proposed Solar Particle Acceleration Radiation and Kinetics \citep{2023Aeros..10.1034R} mission concept, will play a crucial role in advancing our understanding of QPPs and small-scale energy release processes in the solar atmosphere. In addition, coordinated observations \citep{2025A&A...701A..77B} between Solar Orbiter and Daniel K. Inouye Solar Telescope \citep{2020SoPh..295..172R} will provide great opportunities for this.

\begin{acknowledgements}
      Solar Orbiter is a space mission of international collaboration between ESA and NASA, operated by ESA. The EUI instrument was built by CSL, IAS, MPS, MSSL/UCL, PMOD/WRC, ROB, LCF/IO with funding from the Belgian Federal Science Policy Office (BELSPO/PRODEX PEA C4000134088, 4000112292 and 4000106864); the Centre National d’Etudes Spatiales (CNES); the UK Space Agency (UKSA); the Bundesministerium für Wirtschaft und Energie (BMWi) through the Deutsches Zentrum für Luft- und Raumfahrt (DLR); and the Swiss Space Office (SSO). The research that led to these results was subsidised by the Belgian Federal Science Policy Office through the contract B2/223/P1/CLOSE-UP. DL was supported by a Senior Research Project (G088021N) of the FWO Vlaanderen. TVD was supported by the C1 grant TRACEspace of Internal Funds KU Leuven and a Senior Research Project (G088021N) of the FWO Vlaanderen. Furthermore, TVD received financial support from the Flemish Government under the long-term structural Methusalem funding program, project SOUL: Stellar evolution in full glory, grant METH/24/012 at KU Leuven. The paper is also part of the DynaSun project and has thus received funding under the Horizon Europe programme of the European Union under grant agreement (no. 101131534). Views and opinions expressed are however those of the author(s) only and do not necessarily reflect those of the European Union and therefore the European Union cannot be held responsible for them. CV thanks the Belgian Federal Science Policy Office (BELSPO) for the provision of financial support in the framework of the PRODEX Programme of the European Space Agency (ESA) under contract numbers 4000143743 and 4000134088. 
\end{acknowledgements}

\bibliographystyle{aa} 
\bibliography{Lim_bib} 

\begin{appendix} 

\section{Examples of EUV brightenings}\label{appendix:example}
\begin{figure*}
  \resizebox{\hsize}{!}{\includegraphics{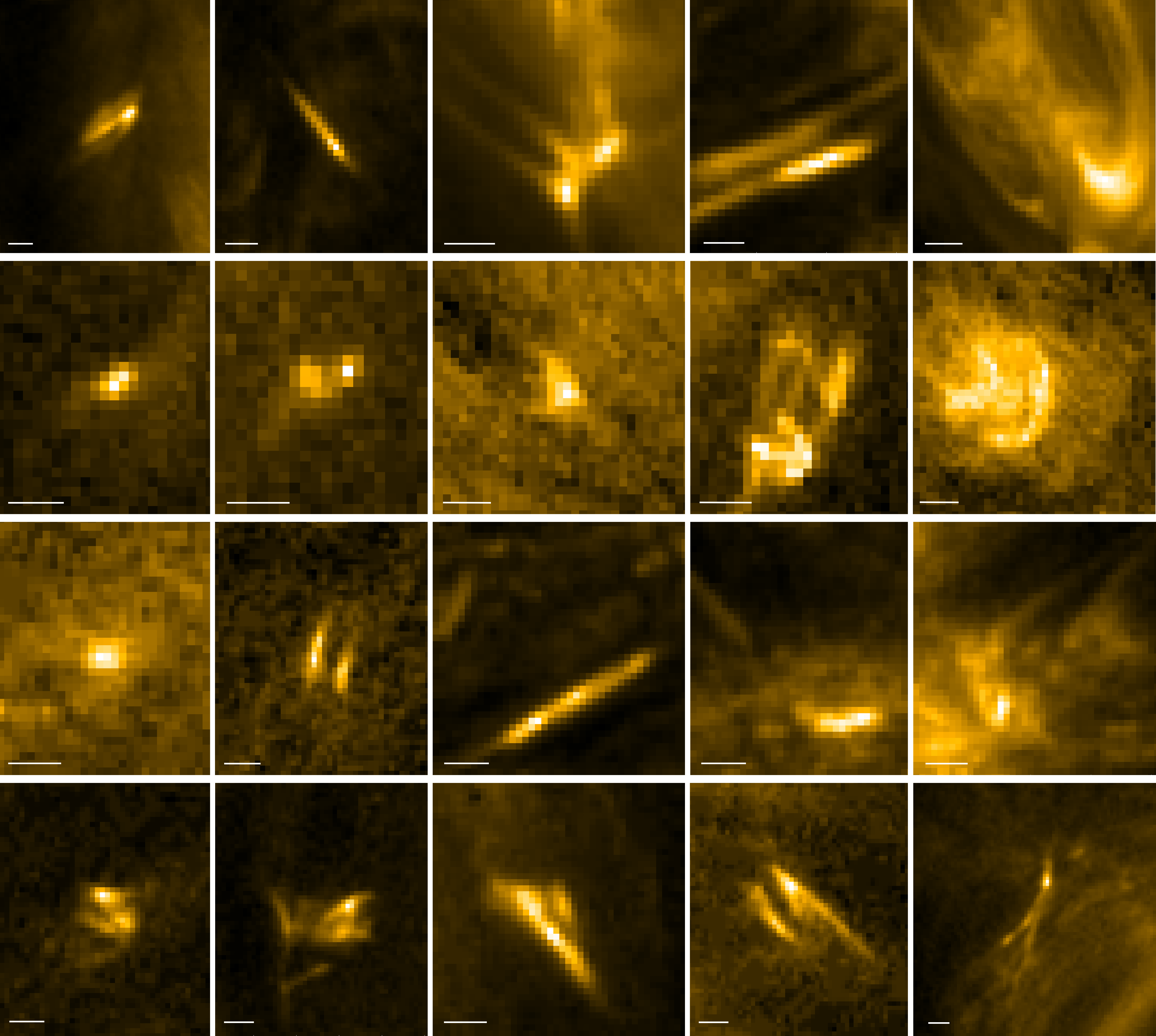}}
  \caption{Additional examples of 20 EUV brightenings, with five events shown from each dataset. Top row: active region on 19 October 2024; second row: quiet Sun on 19 October 2024; third row: active region on 19 March 2025; bottom row: quiet Sun on 19 March 2025. The white solid line denotes 1~Mm.}
  \label{fig:appendix_campfires}
\end{figure*}

Fig.~\ref{fig:appendix_campfires} shows additional examples of the detected EUV brightenings from each dataset.

\section{Threshold Sensitivity Test}\label{appendix:influence_threshold}

\begin{figure*}
  \resizebox{\hsize}{!}{\includegraphics{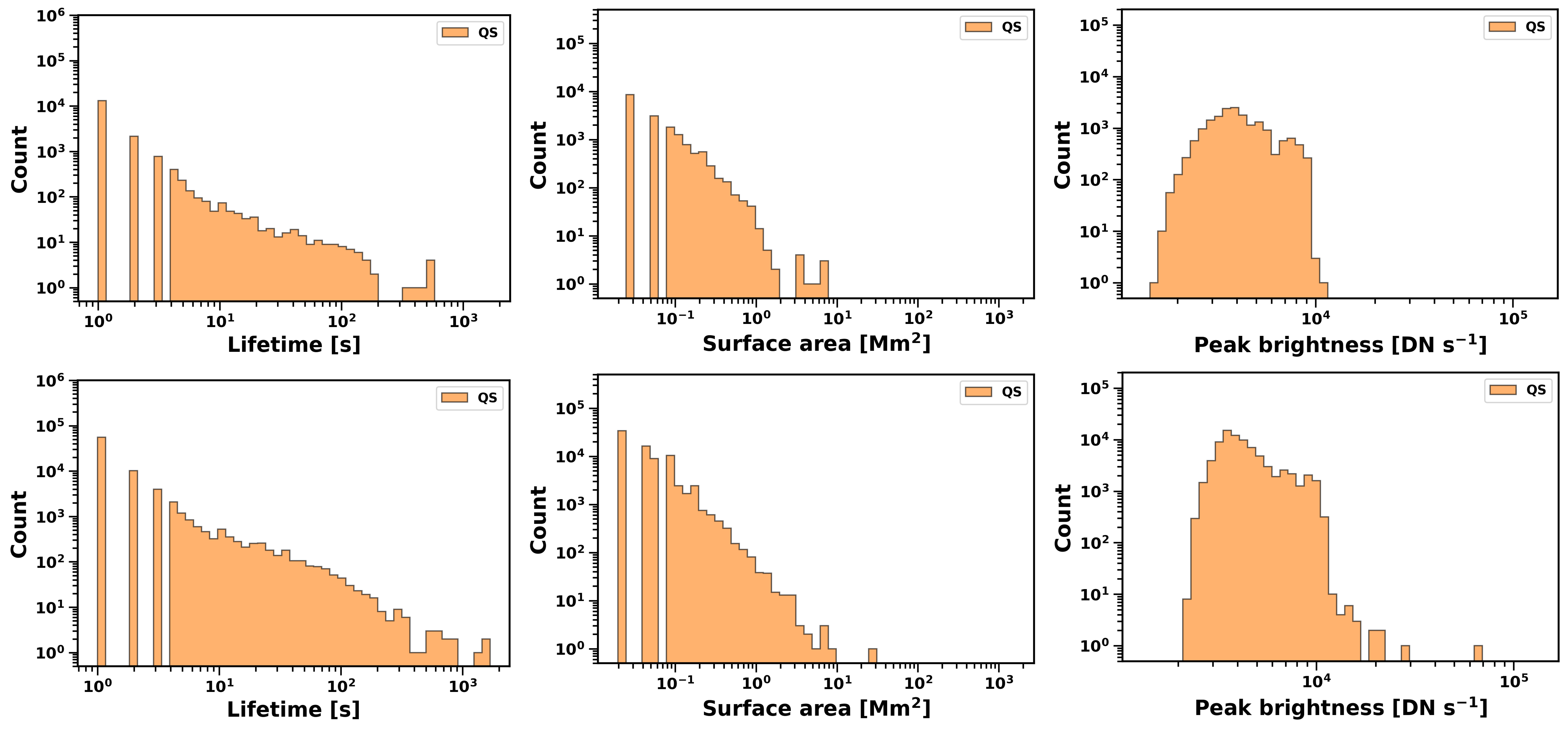}}
  \caption{Logarithmic histograms of the lifetime (left), surface area (middle), and peak brightness (right) of EUV brightenings detected in quiet Sun regions on 19 October 2024 (top) and 19 March 2025 (bottom). The detection scheme was applied with a threshold $n=6$ for both datasets.}
  \label{fig:appendix_threshold}
\end{figure*}

To examine the sensitivity of our detection results to the choice of threshold, we compared the parameter histograms obtained with different values of thresholds. Fig.~\ref{fig:appendix_threshold} shows the histrograms of lifetime, surface area, and peak brightness for the QS regions when a threshold of $n=6$ was applied. As expected, the total number of detected events decreases for higher thresholds. However, the overall shapes of the distributions remain nearly unchanged, indicating that the statistical properties derived from the detected events, such as extrema, average values, and correlation coefficients between parameters, are largely insensitive to the exact choice of threshold.

\section{Influene of Temporal Cadence on Detection Results}\label{appendix:influence_cadence}

To assess the influence of temporal resolution on the detection of short-lived events, we rebinned the original 1~s cadence dataset of the QS region on 19 October 2024 to emulate cadences of 3~s and 10~s (resulting in effective exposure times of approximately 1.98~s and 6.59~s) by averaging every three and ten consecutive frames, respectively. The same detection algorithm and threshold were applied, and the resulting histograms were constructed using identical binning for all cases to ensure direct comparability. 

\begin{figure*}
  \resizebox{\hsize}{!}{\includegraphics{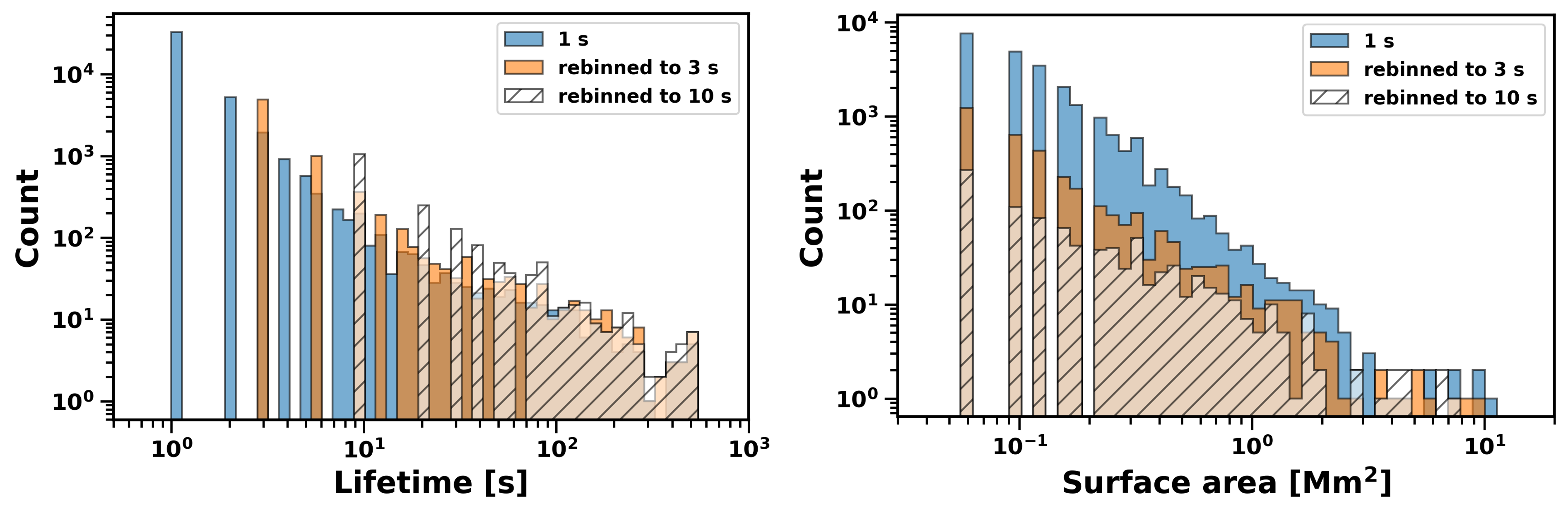}}
  \caption{Histograms of lifetimes (left) and surface areas (right) of EUV brightenings detected in the QS region on 19 October 2024. The results from the original 1~s cadence data (blue solid bars) are compared with datasets rebinned to 3~s (orange solid bars) and 10~s (hatched bars) cadence.}
  \label{fig:appendix_cadence}
\end{figure*}

Fig.~\ref{fig:appendix_cadence} shows that the minimum detectable lifetime is strongly cadence dependent. Moreover, the total number of detected events increases as cadence is reduced, because multiple short-lived EUV brightenings become blended into longer-lived ones, thereby shifting the overall distribution toward longer lifetimes. The surface area distributions exhibit a similar systematic effect. The slope decreases with lower cadence, reflecting a systematic reduction in the detection of smaller events. These results demonstrate that reduced temporal resolution biases both the lifetime and surface area distributions, favouring the detection of longer-lived and larger events at the expense of shorter-lived and smaller ones.

\end{appendix}

\end{document}